\documentclass[12pt]{article}
\usepackage{ascmac}
\usepackage{ulem}
\usepackage{latexsym}
\usepackage{graphicx}
\usepackage{amsmath}
\usepackage[top=3.5cm,bottom=3.5cm,left=2cm,right=2cm]{geometry}
\usepackage{pifont}
\usepackage{cancel}

\newcommand{\h}{\hspace}
\newcommand{\be}{\begin{equation}}
\newcommand{\e}{\end{equation}}
\begin{document}

\title{
\vbox{
\baselineskip 14pt
\hfill \hbox{\normalsize KUNS-2519}
} \vskip 1cm
\bf \Large  Weak Scale From the Maximum Entropy Principle\vskip 0.5cm
}
\author{Yuta~Hamada\thanks{E-mail:  \tt hamada@gauge.scphys.kyoto-u.ac.jp},
Hikaru~Kawai\thanks{E-mail: \tt hkawai@gauge.scphys.kyoto-u.ac.jp}, and 
Kiyoharu~Kawana\thanks{E-mail: \tt kiyokawa@gauge.scphys.kyoto-u.ac.jp}
\bigskip\\
\it \normalsize
 Department of Physics, Kyoto University, Kyoto 606-8502, Japan\\
\smallskip
}
\date{\today}

\maketitle

\abstract{\noindent\normalsize
}The theory of multiverse and wormholes suggests that the parameters of the Standard Model are fixed in such a way that the radiation of the $S^{3}$ universe at the final stage $S_{rad}$ becomes maximum, which we call the maximum entropy principle. Although it is difficult to confirm this principle generally, for a few parameters of the Standard Model, we can check whether $S_{rad}$ actually becomes maximum at the observed values. In this paper, we regard $S_{rad}$ at the final stage as a function of the weak scale ( the Higgs expectation value ) $v_{h}$, and show that it becomes maximum around $v_{h}={\cal{O}}(300\text{GeV})$ when the dimensionless couplings in the Standard Model, that is, the Higgs self coupling, the gauge couplings, and the Yukawa couplings are fixed. Roughly speaking, we find that the weak scale is given by
\begin{equation} v_{h}\sim\frac{T_{BBN}^{2}}{M_{pl}y_{e}^{5}},\nonumber\end{equation}
where $y_{e}$ is the Yukawa coupling of electron, $T_{BBN}$ is the temperature where the Big Bang Nucleosynthesis starts and $M_{pl}$ is the Planck mass.

\newpage
\begin{figure}[!h]
\begin{center}
\includegraphics[width=11cm]{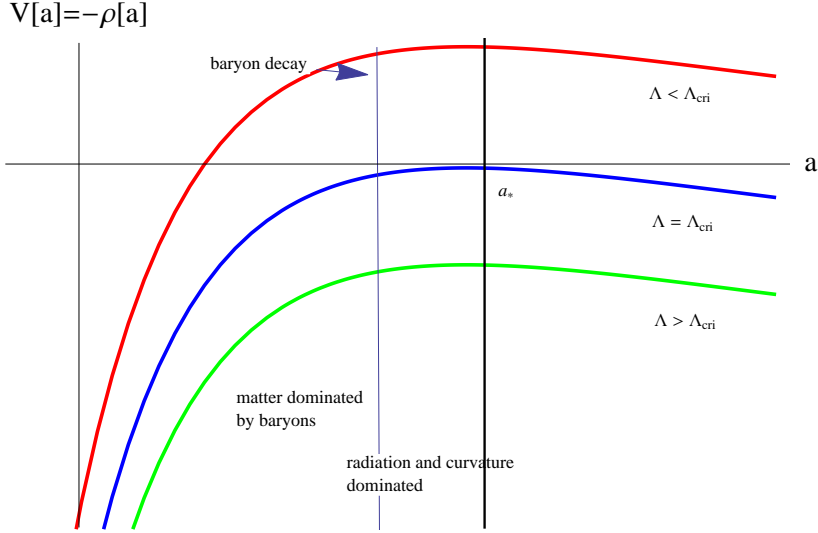}
\end{center}
\caption{The potential of the $S^{3}$ universe. If the Cosmological Constant $\Lambda$ is chosen so that the maximum of $V(a)$ becomes zero, the universe takes infinite time to grow up to the size $a_{*}$. The blue line is required by the maximum entropy principle.
}
\label{fig:pot}
\end{figure}

\section{Introduction and Review of History of Universe}\label{sec:int}
The theory of multiverse and wormholes \cite{Coleman:1988tj,Kawai:2011rj,Kawai:2011qb,Kawai:2013wwa,Hamada:2014ofa} suggests that the parameters of our universe are fixed in such a way that the radiation of the universe $S_{rad}$ at the final stage becomes maximum, which we call the maximum entropy principle or the Big Fix. Here, "the final stage" means that the curvature term balances with the other contents of the universe, and the radius of the universe $a$ is getting close to the critical value $a_{*}$ spending infinite time. See Fig.\ref{fig:pot} for example and see \cite{Kawai:2011rj,Kawai:2011qb,Kawai:2013wwa,Hamada:2014ofa} for the details. This assertion can be checked in principle by changing the parameters of the Standard Model (SM) one by one, if we know how $S_{rad}$ is determined by those parameters. In general, this procedure is difficult to do because of the lack of our understanding of the history of the universe. However for a few couplings of the SM, we can estimate their effects on $S_{rad}$ under some assumptions, and we can actually check the principle. As a concrete example we consider the Higgs expectation value $v_{h}$. Although there are many possibilities for the final stage of the universe, we assume the following scenario: \\\
{\underline{\it{Assumptions for the Final Stage of the Universe}}}\\
\\
(I): The Dark Matter (DM) decays much earlier than the baryons. This guarantees that the radiation produced by the DM is negligible compared with that of baryons.\\
\\
(II): The Cosmological Constant (CC) at the final stage of the universe is fixed to the critical value $\Lambda_{cri}$ so that the curvature term balances with the radiation produced by the baryon decay (see Fig.\ref{fig:pot}). The maximum entropy principle predicts that the dark energy should decrease from the present value to $\Lambda_{cr}$ in the future.  
\\
\\
(III): Baryons decay with the life time $\tau_{B}$, and the radiation $S_{rad}$ is produced. In this paper, we assume that
\be \tau_{B}=10^{36}\h{2mm}\text{year}.\e
After that, the radiation and the CC balance with the curvature term while electrons, positrons and neutrinos annihilate into photons. \\
\\
\\
Based on these assumptions, we show an evidence of the Big Fix: When the Higgs self coupling, gauge couplings and Yukawa couplings are fixed, $S_{rad}$ becomes maximum around $v_{h}={\cal{O}}(300\text{GeV})$\footnote{In the previous paper \cite{Hamada:2014ofa}, we have shown that $S_{rad}$ becomes maximum around $v_{h}={\cal{O}}(300\text{GeV})$ when the Higgs self coupling, the gauge couplings and the current quark and lepton masses are fixed.}.
We first review how $S_{rad}$ is produced through the history of the universe. In the following argument, we denote the photon temperature by $T$.\\
\begin{itemize}\item {\it{Stage 1}}: The baryon number $N_{B}$ is produced by the sphaleron process if we assume the standard leptogenesis scenario. Here we briefly summarize how $N_{B}$ is produced in the early universe. The number density of a particle species $i$ is given by 
\begin{align} n_{i}&=\frac{g_{i}}{\left(2\pi\right)^{3}}\int\frac{dp^{3}}{\exp\left(\frac{\sqrt{p^{2}+m_{i}^{2}}-\mu_{i}}{T}\right)\pm1},\label{density1}
\end{align}
where $g_{i}$ is the degree of freedom, $\mu_{i}$ is the chemical potential and the sign $\pm$ is $-$ for boson and $+$ for fermion. Then the difference between a particle and an anti-particle is given by
\begin{align} n_{i}-\bar{n}_{i}
&:=g_{i}\mu_{i}\frac{T^{2}}{6}\begin{cases}K_{f}(\frac{m_{i}}{T})&\text{(for fermion)},\\
\\
K_{b}(\frac{m_{i}}{T})&\text{(for boson)},\end{cases}\end{align}
where
\be K_{f}(y)=\frac{3}{2\pi^{2}}\int_{0}^{\infty}dx\h{1mm}x^{2}\cosh^{-2}\left(\frac{\sqrt{x^{2}+y^{2}}}{2}\right)\h{2mm},\h{2mm}K_{b}(y)=\frac{3}{2\pi^{2}}\int_{0}^{\infty}dx\h{1mm}x^{2}\sinh^{-2}\left(\frac{\sqrt{x^{2}+y^{2}}}{2}\right).\e
We can eliminate the chemical potentials $\{\mu_{i}\}$ by using the conservations of the total electromagnetic charge $Q$ and the difference between the baryon number and the lepton number $N_{B}-N_{L}$. As a result, the baryon number $N_{B}$ is given as a function of  $\frac{m_{i}}{T}$:
\be N_{B}=N_{B}\left(\frac{m_{i}}{T}\right).\e
See Appendix B in \cite{Hamada:2014ofa} for the explicit formula and detailed calculations. When $T$ reaches the sphaleron decoupling temperature $T_{\text{sph}}$, $N_{B}$ is fixed to $N_{B}(\frac{m_{i}}{T_{sph}})$. We can obtain $T_{\text{sph}}$ by using the recent numerical result \cite{D'Onofrio:2012ni}:
\be T_{\text{sph}}=\frac{7}{8}\times\frac{160v_{h}}{246}\text{GeV}.\e
Here, we have assumed that the Higgs self coupling and the gauge couplings are fixed. Thus, $N_{B}$ at $T_{sph}$ is given by
\be N_{B}=N_{B}\left(\frac{8m_{i}}{7v_{h}}\times\frac{246}{160}\right).\label{eq:fixedNB}\e
Note that if we fix the gauge couplings and the Yukawa couplings, $N_{B}$ is just a constant because the quark masses $m_{q}$ and the gauge boson mass $m_{W}$ are given by 
\be  m_{q}=\frac{y_{q}v_{h}}{\sqrt{2}}\h{2.5mm},\h{2.5mm}m_{W}=\frac{g_{2}v_{h}}{2}.\e
\end{itemize}
\begin{itemize}\item {\it{Stage 2}}: The ratio of neutrons to all nucleons $X_{n}$ is fixed by the following processes:\\
  1)For $T>1$MeV, protons and neutrons are in thermal equilibrium through the weak interactions, and $X_{n}$ at that time is given by
\be X_{n}=\frac{1}{1+\exp(\frac{Q}{T})},\e
where $Q:=m_{n}-m_{p}$ is the mass difference between a neutron and a proton. \\
\\
  2)The weak interactions are frozen out, and $X_{n}$ decreases through the beta decay until $T$ reaches the temperature $T_{BBN}$ where the Big Bang Nucleosynthesis (BBN) starts. Here, note that the life time of a neutron $\tau_{n}$ depends strongly on $v_{h}$, $m_{e}$ and $Q-m_{e}$ where $m_{e}$ is the electron mass. After $T_{BBN}$, neutrons are rapidly converted to atomic nuclei. We discuss those processes in more detail in subsection \ref{subsec:Xn}. See also \cite{Hamada:2014ofa}. \end{itemize}
\begin{itemize}\item {\it{Stage 3}}: The radiation at the early universe becomes dilute, and the matter dominated era starts. After that, the dark energy becomes dominant. This is the era where we live. As discussed before, we assume that the dark energy becomes very small in the future. \end{itemize}
\begin{itemize}\item {\it{Stage 4}}: The DM decays sufficiently earlier than baryons.\end{itemize}
\begin{itemize}\item {\it{Stage 5}}: Baryons decay, 
and the radiation $S_{rad}$ is produced. $S_{rad}$ depends on $X_{n}$, the masses of proton and helium nucleus $m_{p},m_{m_{He}}$ and their life times $\tau_{p},\tau_{He}$. Moreover, we must take into account the possibility that a pion produced by the decay of a helium nucleus is scattered by the remaining nucleons, and loses its energy. We denote the energy loss through this process by a dimensionless parameter $\epsilon$.\end{itemize}

\noindent  From the above argument, it is clear that we need to know the $v_{h}$ dependence of the following quantities in order to evaluate $S_{rad}$ as a function of $v_{h}$:
\be m_{p}\h{2mm},\h{2mm}m_{\text{He}}\h{2mm},\h{2mm}Q(\h{1mm}\text{or}\h{1mm}m_{n})\h{2mm},\h{2mm}X_{n}\h{2mm},\h{2mm}T_{BBN}\h{2mm},\h{2mm}\tau_{p}\h{2mm},\h{2mm}\tau_{He}\h{2mm},\h{2mm}\epsilon.\e
In this paper, we use the phenomenological equations for $m_{p}$, $Q$ and $\tau_{p}$:
\be m_{p}=\alpha\Lambda_{QCD}+\beta(2m_{u}+m_{d})\h{2.5mm},\h{2.5mm}Q=\beta(m_{d}-m_{u})-M_{em}\h{2.5mm},\h{2.5mm}m_{\pi}^{2}=\gamma\Lambda_{QCD}\frac{m_{u}+m_{d}}{2},\label{eq:hadronmass}\e
\be  \Gamma_{p}=\tau_{p}^{-1}\propto \frac{m_{p}^{5}}{M_{G}^{4}}G.\label{eq:proton life}
\e
The meanings of the parameters and the factor $G$ are as follows.
\begin{itemize}\item{}$m_{u}$ and $m_{d}$ are masses of an up quark and a down quark. Their typical values are $(m_{u},m_{d})=(2.3\text{MeV},4.8\text{MeV})$ \cite{Beringer:1900zz}.\end{itemize}
\begin{itemize}\item{}$\Lambda_{QCD}$ is the scale where the QCD coupling becomes ${\cal{O}}(1)$, which we fix to $300$MeV in this paper.\end{itemize}
\begin{itemize}\item{}$M_{em}$ is the electromagnetic energy of a neutron, and $\alpha,\beta,\gamma$ are numerical constants that should be determined by QCD. In principle, they can also be found from observations. The typical values we use in this paper are
\be \alpha=3.1\h{2mm},\h{2mm}\beta=1.4\h{2mm},\h{2mm}\gamma=16\h{2mm},\h{2mm}M_{em}=2.2\text{MeV},\label{eq:typicalval}\e
which explain the experimental results \cite{Beringer:1900zz} well.
\end{itemize}
\begin{itemize}\item{}$G$ can be approximately calculated by using the effective interaction of the proton decay as 
\be G=constant\times\left(1-\left(\frac{m_{\pi}}{m_{p}}\right)^{2}\right)^{2}.\label{eq:effG}\e
Here, we neglect the electron mass $m_{e}$ in the formula for the proton life time (see \ref{app:proton}). 
In the limit $m_{u,d}\ll m_{p}$, we expect that $G$ depends linearly on $\frac{m_{u,d}}{m_{p}}$:
\be G\simeq1+\xi\frac{m_{u,d}}{m_{p}}\h{3mm}(\text{for $m_{u,d}\ll m_{p}$}).\label{eq:gfunc}\e
 \end{itemize}
On the other hand, for $m_{He}$, $\tau_{He}$ and $\epsilon$, it is difficult to determine their $v_{h}$ dependence because of the complicated effects of nuclear physics. However, by the numerical calculations (see Section\ref{sec:radiation} and also \cite{Hamada:2014ofa}), we can show that these quantities effectively contribute to $S_{rad}$ through a single function $c\left(\epsilon,\frac{\tau_{He}}{\tau_{p}},\frac{m_{He}}{m_{p}}\right)$ whose typical value is of ${\cal{O}}(0.01)$. 
In this paper, we assume that $c$ is a constant, namely, we neglect the $v_{h}$ dependence of $m_{He}$, $\tau_{He}$ and $\epsilon$. The details are given in Section\ref{sec:radiation} and Section\ref{sec:summary}.  
Finally, we also assume that $T_{BBN}$ does not depend on $v_{h}$ in the main part of this paper, and fix it to $0.1$MeV. The case where it depends on $v_{h}$ is discussed in \ref{app:bd}.\\
\\
{\it{\textbf{Summary}}} : Assuming that
\be \alpha\h{2mm},\h{2mm}\beta\h{2mm},\h{2mm}\gamma\h{2mm},\h{2mm}c\left(\epsilon,\frac{\tau_{He}}{\tau_{p}},\frac{m_{He}}{m_{p}}\right)\h{2mm},\h{2mm}M_{em}\label{eq:unknown}\e
are fixed to the phenomenologically reasonable values, we find that $S_{rad}$ has a global maximum around $v_{h}={\cal{O}}(300\text{GeV})$ when the Higgs self coupling, the gauge couplings and the Yukawa couplings are fixed. Here, we consider such a region that a neutron is heavier than a proton. \\

This paper is organized as follows. In Section\ref{sec:radiation}, we briefly review how the radiation of the universe is determined, and give a qualitative expression. In Section\ref{sec:para}, we discuss the $v_{h}$ dependence of $S_{rad}$. In Section\ref{sec:big}, we consider the Big Fix. In Section\ref{sec:summary}, we give summary and discussion.

\section{Radiation of the Universe at the Final Stage}\label{sec:radiation}
The radiation of the universe at the final stage can be obtained in principle by solving the following equations:
\begin{align} \frac{dN_{p}(t)}{dt}&=-\tau_{p}^{-1}\cdot N_{p}(t)+3\tau^{-1}_{He}\cdot N_{He}(t),\label{eq:Npdiff}\\
 \frac{dN_{He}(t)}{dt}&=-\tau_{He}^{-1}\cdot N_{He}(t),\label{eq:NHediff}\\
H^{2}(t):&=\left(\frac{\dot{a}}{a}\right)^{2}=\frac{1}{3M_{pl}^{2}}\cdot\left(\frac{M(t)}{a^{3}}+\frac{S_{rad}(t)}{a^{4}}-\frac{M_{pl}^{2}}{a^{2}}+M_{pl}^{2}\Lambda\right),\label{friedman}\\
 M(t)&=m_{p}\cdot N_{p}(t)+m_{He}\cdot N_{He}(t),
\\
 \frac{dS_{rad}(t)}{dt}&=a(t)m_{p}\times\left(\tau_{p}^{-1}\cdot N_{p}(t)+(1-2\epsilon)\cdot\tau_{He}^{-1}\cdot N_{He}(t)\right).\label{eq: radiation2}\end{align}
Here $N_{p}(t)$ and $N_{\text{He}}(t)$ are the numbers of protons and helium nuclei, and $\epsilon$ represents the following effect: a pion produced by the nucleon decay of a helium nucleus is scattered by the remaining nucleons, loses its kinetic energy and produces less radiation after it decays \cite{Hamada:2014ofa}. The initial values of $N_{p}(t),N_{He}(t)$ are given by
\be N_{p}(0)=N_{B}(1-2X_{n}),\e
\be N_{He}(0)=N_{B}\frac{X_{n}}{2},\e
where $N_{B}$ is the total baryon number and $X_{n}$ is the ratio of neutrons to all nucleons. By numerical calculations, we have obtained the qualitative expression of $S_{rad}$ \cite{Hamada:2014ofa}:
\be S_{rad}= constant \times\left(\frac{1}{M_{pl}^{2}}\right)^{\frac{1}{3}}\times (N_{B}m_{p})^{\frac{4}{3}}\tau^{\frac{2}{3}}_{p} \times\{1-c\left(\epsilon,\frac{\tau_{He}}{\tau_{p}},\frac{m_{He}}{m_{p}}\right)X_{n}\}.\label{eq:apprad}\e
Qualitative meaning of this equation is as follows: First, if there is no atomic nuclei, baryons are all protons having its life time $\tau_{p}$. If we simplify the situation so that these protons decay similitaneously, we can obtain the radiation $S_{rad}$ by the energy conservation
\be \frac{N_{B}m_{p}}{a^{3}(\tau_{p})}=\frac{ S_{rad}}{a^{4}(\tau_{p})},\e
from which we have
\be S_{rad}=N_{B}m_{p}a(\tau_{p}).\e
In our scenario, because the universe is matter dominated until $\tau_{p}$, the Friedman equation indicates
\be H^{2}(\tau_{p})\simeq\frac{1}{\tau^{2}_{p}}\simeq\frac{N_{B}m_{p}}{M_{pl}^{2}a^{3}(\tau_{p})}.\e
Thus, we obtain
\be S_{rad}\simeq (N_{B}m_{p})^{\frac{4}{3}}\tau^{\frac{2}{3}}_{p}.\e
This expression is modified by the existence of the atomic nuclei. The $1-c\cdot X_{n}$ term in Eq.(\ref{eq:apprad}) represents the such effects. If $c$ is positive, the radiation decreases and vice versa. 

As discussed in Section\ref{sec:int}, $N_{B}$ and $X_{n}$ depend on the SM parameters as
\be N_{B}=N_{B}(\frac{m_{i}}{v_{h}}),\label{eq:baryon}\e
\be X_{n}=X_{n}(v_{h},m_{e},Q,T_{\text{BBN}}).\e
If the Higgs self coupling, the gauge couplings and the Yukawa couplings are fixed\footnote{On the other hands, if the Higgs self coupling, the gauge couplings and the current quark masses are fixed, the $v_{h}$ dependence of $S_{rad}$ comes from $X_{n}$ and $N_{B}$:
\be S_{rad}|_{\text{mass fix}}\propto constant\times (1-cX_{n}(v_{h}))\times N_{B}^{\frac{4}{3}}(v_{h}).\nonumber\label{eq:radiationcurrent}\e
This is because $m_{p}$, $m_{\text{He}}$, $\tau_{p}$, $\tau_{He}$ and $\epsilon$ are determined once $\Lambda_{QCD}$ and the current quark masses are fixed. The conclusion in the previous paper \cite{Hamada:2014ofa} is that $S_{rad}|_{\text{mass fix}}$ becomes maximum around $v_{h}=246$GeV.}, the $v_{h}$ dependence of $S_{rad}$ comes from $m_{p}$, $\tau_{p}$ and $X_{n}$ in Eq.(\ref{eq:apprad}):
\be S_{rad}= constant\times m_{p}^{\frac{4}{3}}(v_{h})\tau_{p}^{\frac{2}{3}}(v_{h})(1-cX_{n}(v_{h})).\label{eq:radiationyukawa}\e
Here, note that because the baryon number $N_{B}$ depends on the SM parameters through $\frac{m_{i}}{v_{h}}$ (see Eq.(\ref{eq:fixedNB})), it is just a constant in this case. While the analytic expressions of $m_{p}$, $m_{\pi}$ and $\tau_{p}$ are given by Eq.(\ref{eq:hadronmass}) and (\ref{eq:proton life}), it is difficult to give the counterparts to $m_{\text{He}},\tau_{\text{He}}$ and $\epsilon$. However, because $c$ varies by only a few percent for $v_{h}<1$TeV \footnote{This can be understood intuitively; because $m_{\text{He}}$, $\tau_{\text{He}}$ and $\epsilon$ depend on $v_{h}$ through the current quark masses, the changes of $\frac{m_{He}}{m_{p}}$, $\frac{\tau_{He}}{\tau_{p}}$ and $\epsilon$ are essentially ${\cal{O}}(\frac{m_{u,d}}{m_{p}})\simeq1\h{1mm}\%$ when we vary $v_{h}$ in the region $v_{h}<1$TeV.}, the $v_{h}$ dependence of $c$ is not so important compared with that of $X_{n}$. In this paper, we assume that $c$ is a constant such as $1/50$ or $1/100$ \cite{Hamada:2014ofa}. We discuss this point in \ref{app:c}.

\section{$v_{h}$ Dependence of $S_{rad}$}\label{sec:para}
In this section, we discuss how $S_{rad}$ depends on $v_{h}$. First, we consider $m_{p}^{\frac{4}{3}}\times\tau_{p}^{\frac{2}{3}}$, and then discuss $X_{n}$.
\subsection{$m_{p}^{\frac{4}{3}}\times\tau_{p}^{\frac{2}{3}}$}
\begin{figure}
\begin{center}
\begin{tabular}{c}
\begin{minipage}{0.5\hsize}
\begin{center}
\includegraphics[width=8cm]{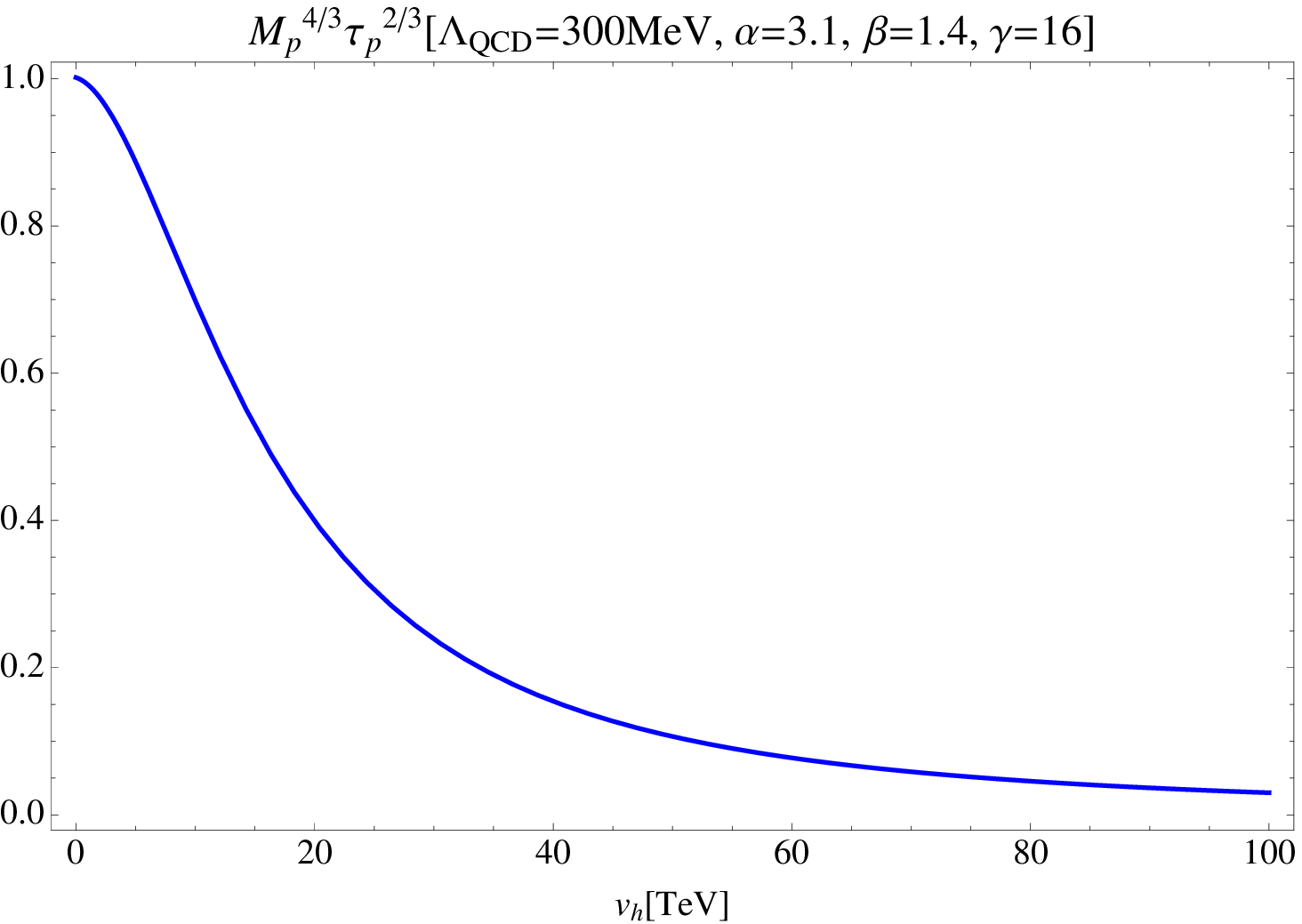}
\end{center}
\end{minipage}
\begin{minipage}{0.5\hsize}
\begin{center}
\includegraphics[width=8cm]{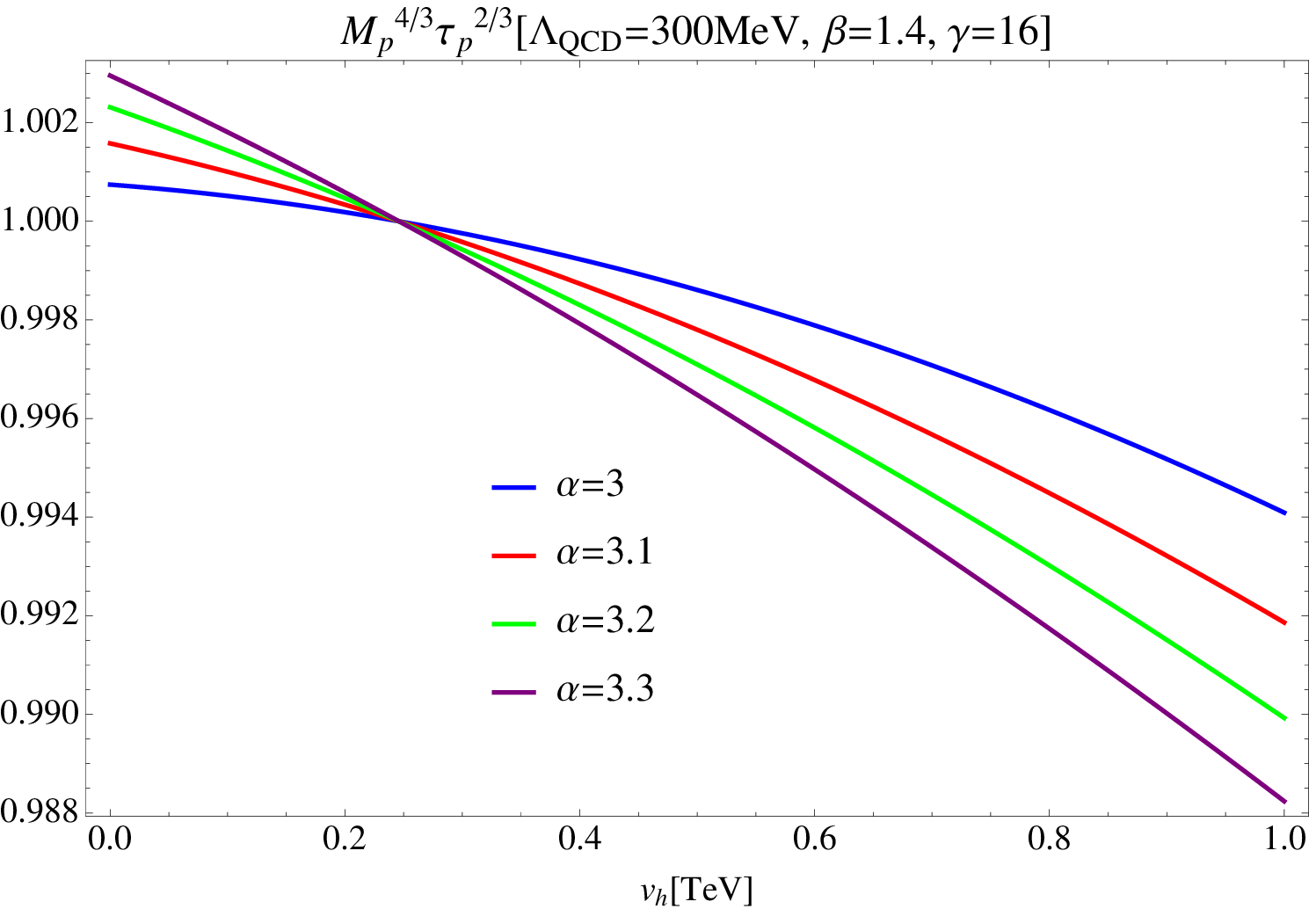}
\end{center}
\end{minipage}
\\
\begin{minipage}{0.5\hsize}
\begin{center}
\includegraphics[width=8cm]{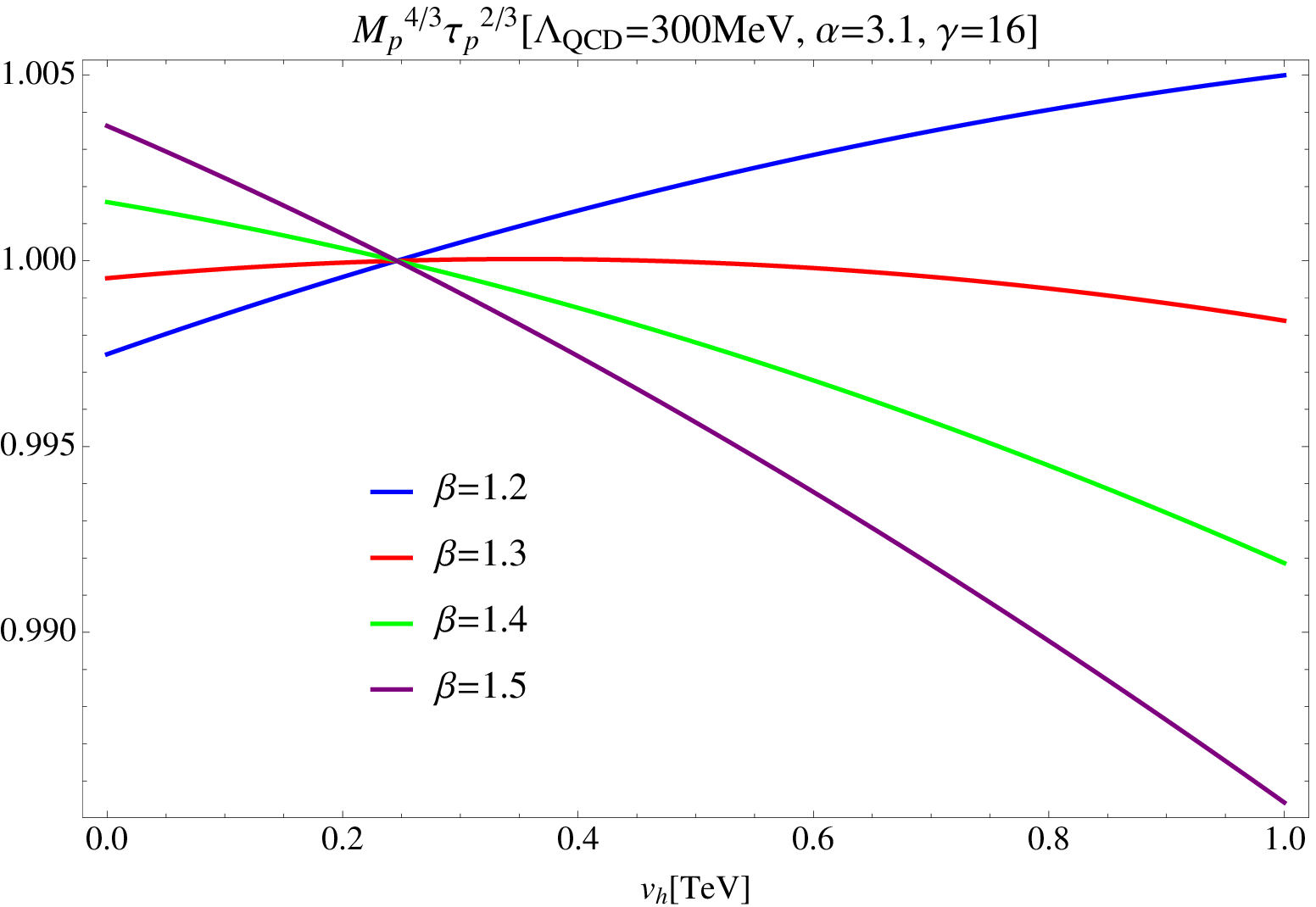}
\end{center}
\end{minipage}
\begin{minipage}{0.5\hsize}
\begin{center}
\includegraphics[width=8cm]{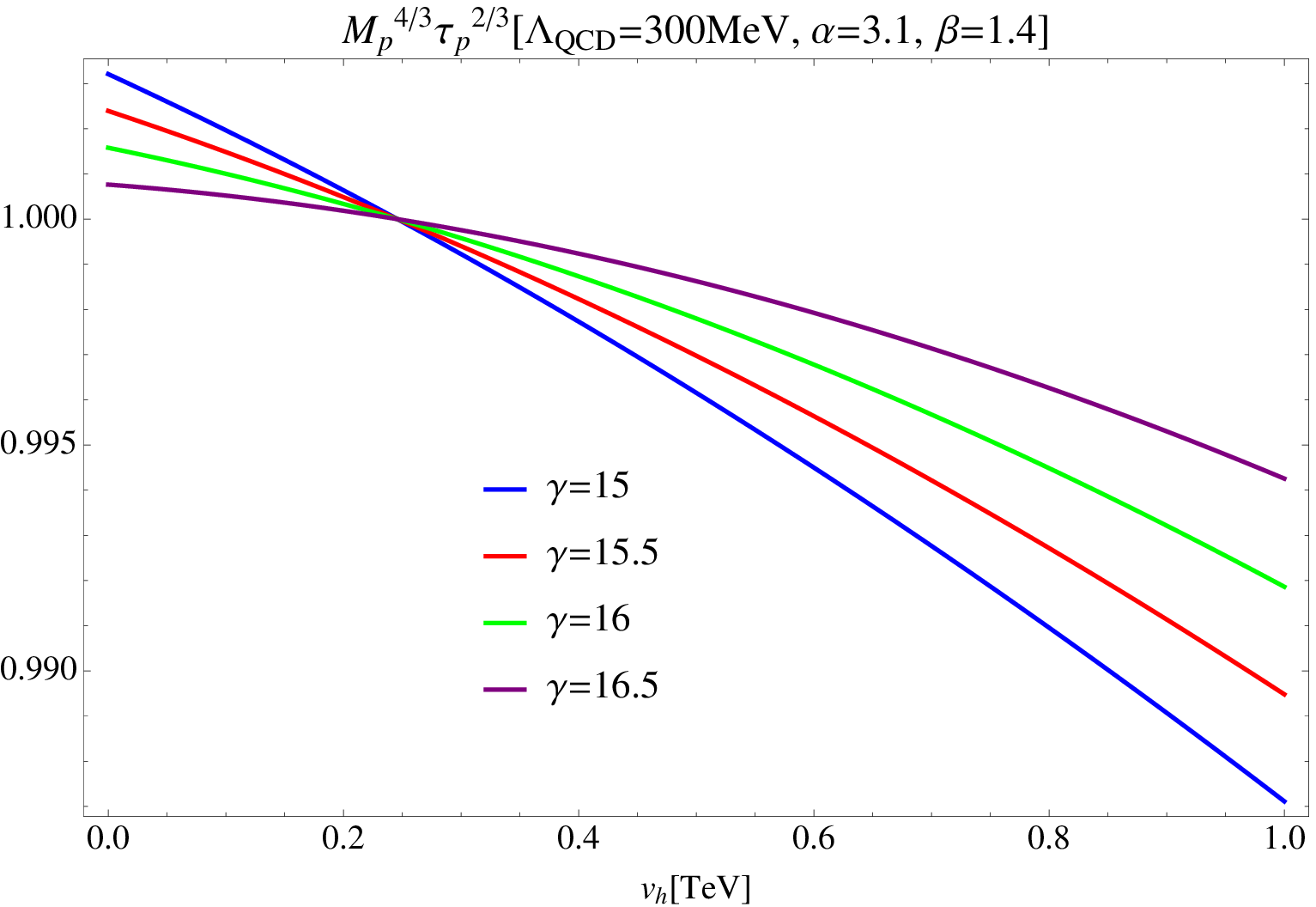}
\end{center}
\end{minipage}
\end{tabular}
\end{center}
\caption{$m_{p}^{\frac{4}{3}}\times\tau_{p}^{\frac{2}{3}}$ as a function of $v_{h}$. Here, the parameters are fixed at the phenomenologically reasonable values, and we have rescaled $m_{p}^{\frac{4}{3}}\times\tau_{p}^{\frac{2}{3}}$ so that it becomes one at $v_{h}=246$GeV. The upper left panel shows the case where $\Lambda_{QCD}=300$MeV, $\alpha=3.1$, $\beta=1.4$ and $\gamma=16$ for $v_{h}<100$TeV. One can see that $m_{p}^{\frac{4}{3}}\times\tau_{p}^{\frac{2}{3}}$ decreases monotonically. The upper right, lower left and lower right panels shows the $v_{h}$ dependence in the region $v_{h}<1$TeV for various values of $\alpha,\beta$ and $\gamma$, respectively.}
\label{fig:masslife}
\end{figure}

In Fig.\ref{fig:masslife}, we show the graphs of 
\be m_{p}^{\frac{4}{3}}\times\tau_{p}^{\frac{2}{3}}\propto\frac{1}{m_{p}^{2}}G^{-\frac{2}{3}}.\e
Here, for $G$ we have used Eq.(\ref{eq:effG}). We can understand Fig.\ref{fig:masslife} qualitatively as follows. In the large $v_{h}$ region, because the current quark masses are larger than $\Lambda_{QCD}$, $G$ becomes a constant, and we have
\be m_{p}^{\frac{4}{3}}\times\tau_{p}^{\frac{2}{3}}\propto\frac{1}{m_{p}^{2}}\simeq\frac{1}{m_{u,d}^{2}}.\e
Therefore, $ m_{p}^{\frac{4}{3}}\times\tau_{p}^{\frac{2}{3}}$ simply decreases in this region. On the other hand, in the $v_{h}<1$TeV region, because the $v_{h}$ dependence appears through the current quark masses $m_{u,d}$, and they are very small compared with $\Lambda_{QCD}$, we can approximate $m_{p}^{\frac{4}{3}}\times\tau_{p}^{\frac{2}{3}}$ lineally as a function of $m_{u,d}$:
\be m_{p}^{\frac{4}{3}}\times\tau_{p}^{\frac{2}{3}}\propto\left(1+\eta(\alpha,\beta,\gamma,\xi)\cdot\frac{m_{u,d}}{\Lambda_{QCD}}\right),\label{eq:defeta}\e
where $\eta$ is given by (see \ref{app:proton})
\be \eta=\left(\frac{4\gamma}{3\alpha^{2}}-\frac{6\beta}{\alpha}\right).\e
As one can see from Fig.\ref{fig:masslife}, for the typical values of $\alpha,\beta$ and $\gamma$ as in Eq.(\ref{eq:typicalval}), $m_{p}^{\frac{4}{3}}\times\tau_{p}^{\frac{2}{3}}$ decreases monotonically as a function of $v_{h}$
, which is crucial for obtaining the maximum of 
\be S_{rad}\propto m_{p}^{\frac{4}{3}}\times\tau_{p}^{\frac{2}{3}}\times\left(1-cX_{n}\right)\e
around $v_{h}={\cal{O}}(300\text{GeV})$.  This is because, as we will see in the next subsection,  $1-cX_{n}$ rapidly increases around $v_{h}={\cal{O}}(200\sim300\text{GeV})$, and becomes almost constant in the $v_{h}>{\cal{O}}(200\sim300\text{GeV})$ region. 

\newpage
\subsection{Neutrons to Nucleons Ratio $X_{n}$}\label{subsec:Xn}
In this subsection, we discuss the $v_{h}$ dependence of $X_{n}$. At a temperature $T\gg1$MeV, neutrons and protons are in thermal equilibrium through the following six processes:
\be n+\nu\leftrightarrow p+e^{-}\h{2mm},\h{2mm}n+e^{+}\leftrightarrow p+\bar{\nu}\h{2mm},\h{2mm}n\leftrightarrow p+e^{-}+\bar{\nu},\e
and $X_{n}$ is given by
\be X_{n}=\frac{1}{1+\exp(\frac{Q}{T})}.\label{eq:equilibrium}\e
The total reaction rate of the $p\rightarrow n$ process is given by \cite{Weinberg}
\be\Gamma(p\rightarrow n)
=0.400\h{1mm}\text{sec$^{-1}$}\times\left(\frac{T}{1\text{MeV}}\right)^{5}\times\left(\frac{246\text{GeV}}{v_{h}}\right)^{4}\times \frac{P\left(\frac{m_{e}}{T},\frac{Q}{T}\right)}{P(0,0)}. \label{eq:reaction rate}\e
Here,
\be P\left(\frac{m_{e}}{T},\frac{Q}{T}\right):=\int_{0}^{\infty}dx\sqrt{1-\left(\frac{m_{e}/T}{Q/T+x}\right)^{2}}\cdot\frac{(Q/T+x)^{2}x^{2}}{\left(1+e^{-xT/T_{\nu}}\right)\left(1+e^{Q/T+x}\right)},\e
and $T_{\nu}$ is the neutrino temperature
\be T_{\nu}=T\times\frac{{\cal{S}}(\frac{m_{e}}{T})}{{\cal{S}}(0)},\e
where 
\be {\cal{S}}(x):=1+\frac{45}{2\pi^{4}}\cdot \int_{0}^{\infty}dy\h{1mm}y^{2}\cdot\left(\sqrt{x^{2}+y^{2}}+\frac{y^{2}}{3\sqrt{x^{2}+y^{2}}}\right)\cdot\frac{1}{1+\exp\left(\sqrt{x^{2}+y^{2}}\right)}.\e
\\

\begin{figure}
\begin{center}
\includegraphics[width=10cm]{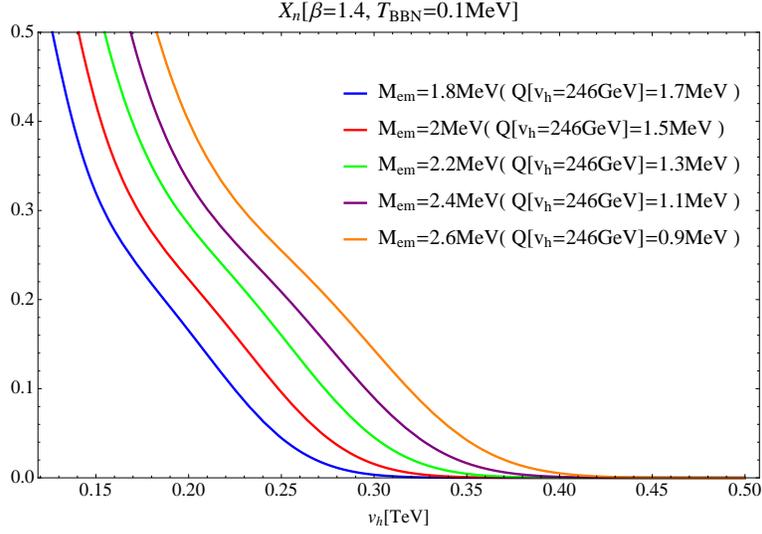}
\caption{ $X_{n}$ as a function of $v_{h}$ when the Yukawa couplings are fixed. Here, $T_{BBN}$ and $\beta$ are fixed to $0.1$MeV and $1.4$, respectively. The lines having the different colors correspond to the different values of $M_{em}$. 
} 
\label{fig:xnvalue}
\end{center}
\end{figure}

As the universe expands, the above processes except for the beta decay decouple at $T_{dec}$. We can obtain $T_{dec}$ by solving
\be H=\Gamma(p\rightarrow n).\e
Below $T_{dec}$, $X_{n}$ decreases through the beta decay until $T$ reaches $T_{BBN}$ where the BBN starts\footnote{ As discussed in Section\ref{sec:int}, we assume that $T_{BBN}$ does not depend on $v_{h}$. The $v_{h}$ dependence of $T_{BBN}$ is discussed in \ref{app:bd}.}. Thus, $X_{n}$ at $T_{BBN}$ is given by
\be X_{n}=\exp\left(-\tau_{n}^{-1}(t_{BBN}-t_{dec})\right)\times X_{n}(t_{dec})=\exp(-\tau_{n}^{-1}(t_{BBN}-t_{dec}))\times\frac{1}{1+\exp(\frac{Q}{T_{dec}})},\label{eq:Xnvalue}\e 
where $\tau_{n}$ is the neutron life time, and the relation between $T$ and $t$ is given by the Friedman equation:
\be H=\frac{1}{2t}=\frac{1}{2}\sqrt{\frac{2\pi^{2} {\cal{N}}}{45M^{2}_{pl}}}T^{2}\label{eq:expansion}.\e
Here, ${\cal{N}}$ is the degrees of freedom and it is given by $43/4$ when $m_{e}<T<m_{\mu}$. By using the Fermi theory, we can calculate $\tau_{n}$ as 
\be \tau_{n}^{-1}=885^{-1}\text{sec}^{-1}\times\left(\frac{m_{e}}{0.51\text{MeV}}\right)^{5}\times\left(\frac{246\text{GeV}}{v_{h}}\right)^{4}\times\frac{F(\frac{Q}{m_{e}})}{F(\frac{1.29}{0.51})},\e
where
\be F(x):=\int_{1}^{x}dy\h{2mm}y(y^{2}-1)^{1/2}(x-y)^{2}.\e
Fig.\ref{fig:xnvalue} shows $X_{n}$ given by Eq.(\ref{eq:Xnvalue}) as a function of $v_{h}$ for various values of $M_{em}$. Here, $T_{BBN}$ is fixed at $0.1$MeV. One can see that $X_{n}$ is a monotonically decreasing function of $v_{h}$, and that $X_{n}$ increases for fixed $v_{h}$ when we increase $M_{em}$. The latter behavior is easily understood; if we increase $M_{em}$, the initial value of $X_{n}$ (which is given by Eq.(\ref{eq:equilibrium})) and the life time of neutron $\tau_{n}$ becomes large because $Q$ becomes small. 

\section{Big Fix of $v_{h}$}\label{sec:big}
\begin{figure}[!h]
\begin{minipage}{0.5\hsize}
\begin{center}
\includegraphics[width=8.5cm]{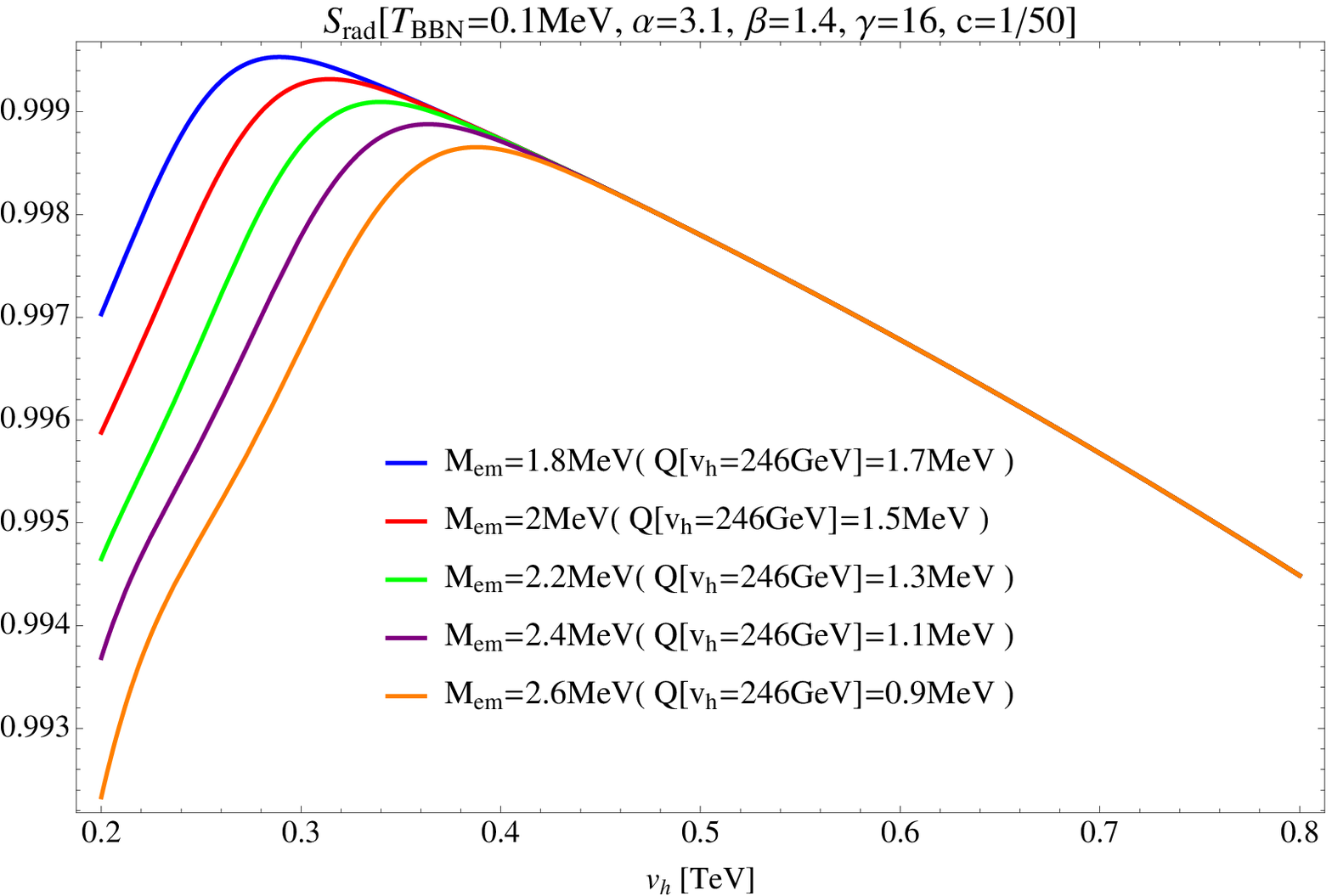}
\end{center}
\end{minipage}
\begin{minipage}{0.5\hsize}
\begin{center}
\includegraphics[width=8.5cm]{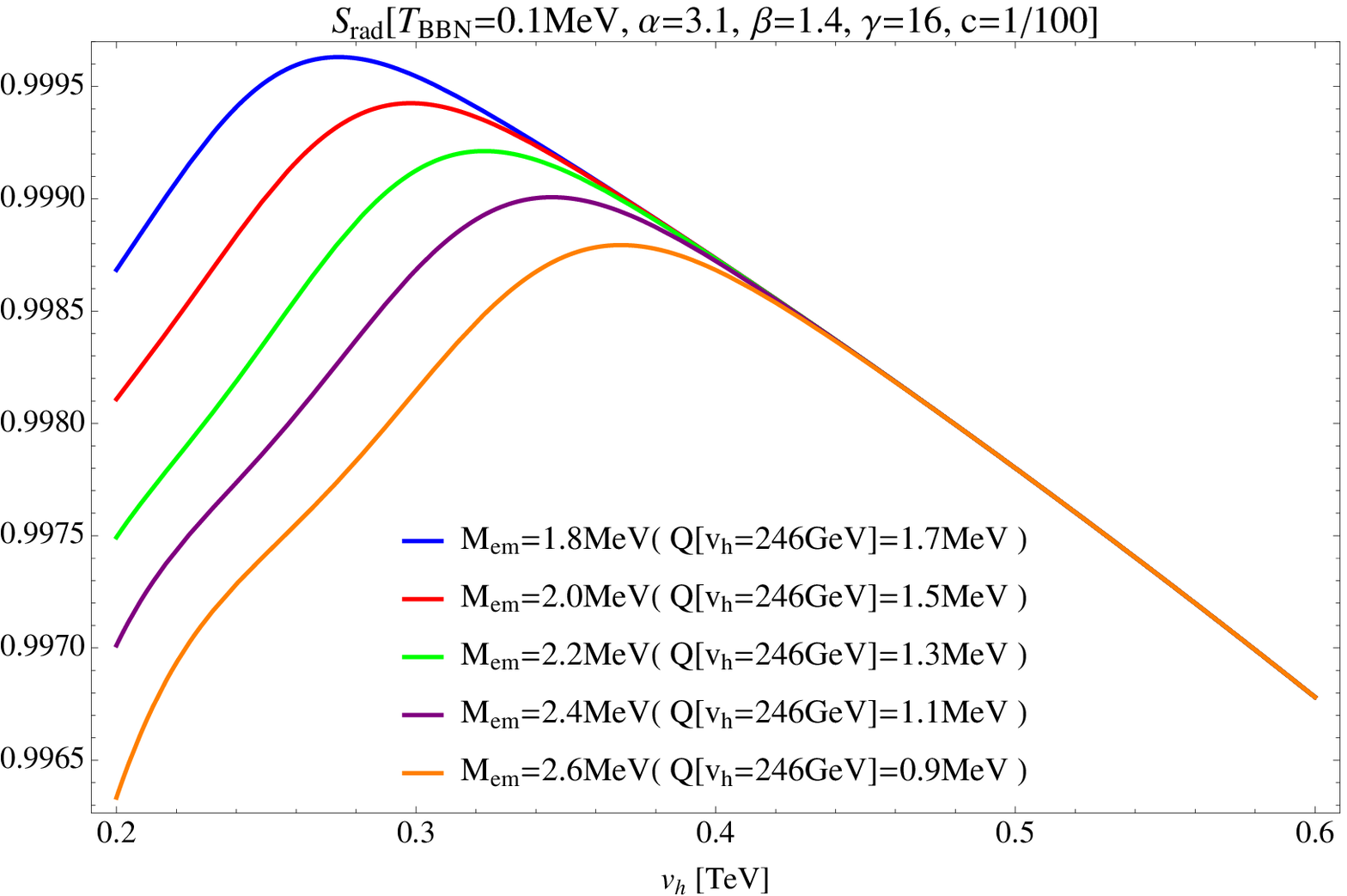}
\end{center}
\end{minipage}
\label{fig:radiation}
\caption{The radiation of the universe $S_{rad}$ as a function of $v_{h}$ for various values of $M_{em}$ when the Higgs self coupling, the gauge couplings and the Yukawa couplings are fixed. Here, $\alpha$, $\beta$ and $\gamma$ are fixed at the typical values, and the scale of the vertical axis is chosen properly. Left (Right) shows the $c=1/50(1/100)$ case. } 
\end{figure}

By using Eq.(\ref{eq:radiationyukawa}) and the results of the previous section, we can determine $S_{rad}$ as a function of $v_{h}$. From Fig.\ref{fig:masslife} and Fig.\ref{fig:xnvalue}, it is clear that $S_{rad}$ has a maximum around $v_{h}={\cal{O}}(300\text{GeV})$. Fig.\ref{fig:radiation} shows the results for the case 
\be T_{BBN}=0.1\text{MeV}\h{2mm},\h{2mm}\alpha=3.1\h{2mm},\h{2mm}\beta=1.4\h{2mm},\h{2mm}\gamma=16.\label{eq:typval}\e
In the left (right) figure, we assume $c=1/50\h{1mm}(1/100)$. $S_{rad}$ has a maximum around $v_{h}={\cal{O}}(300\text{GeV})$. 

\begin{figure}
\begin{center}
\begin{tabular}{c}
\begin{minipage}{0.5\hsize}
\begin{center}
\includegraphics[width=8.5cm]{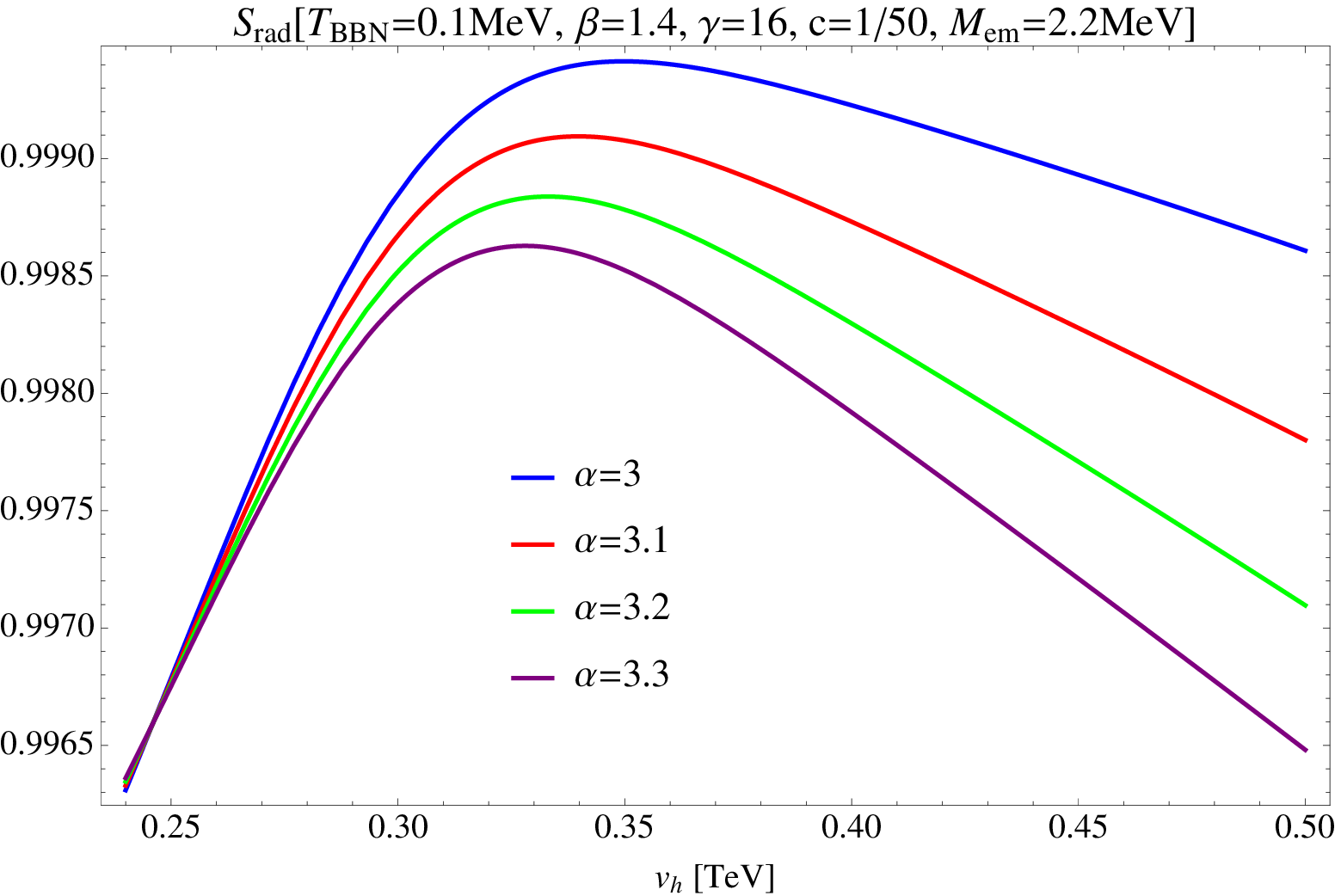}
\end{center}
\end{minipage}
\begin{minipage}{0.5\hsize}
\begin{center}
\includegraphics[width=8.5cm]{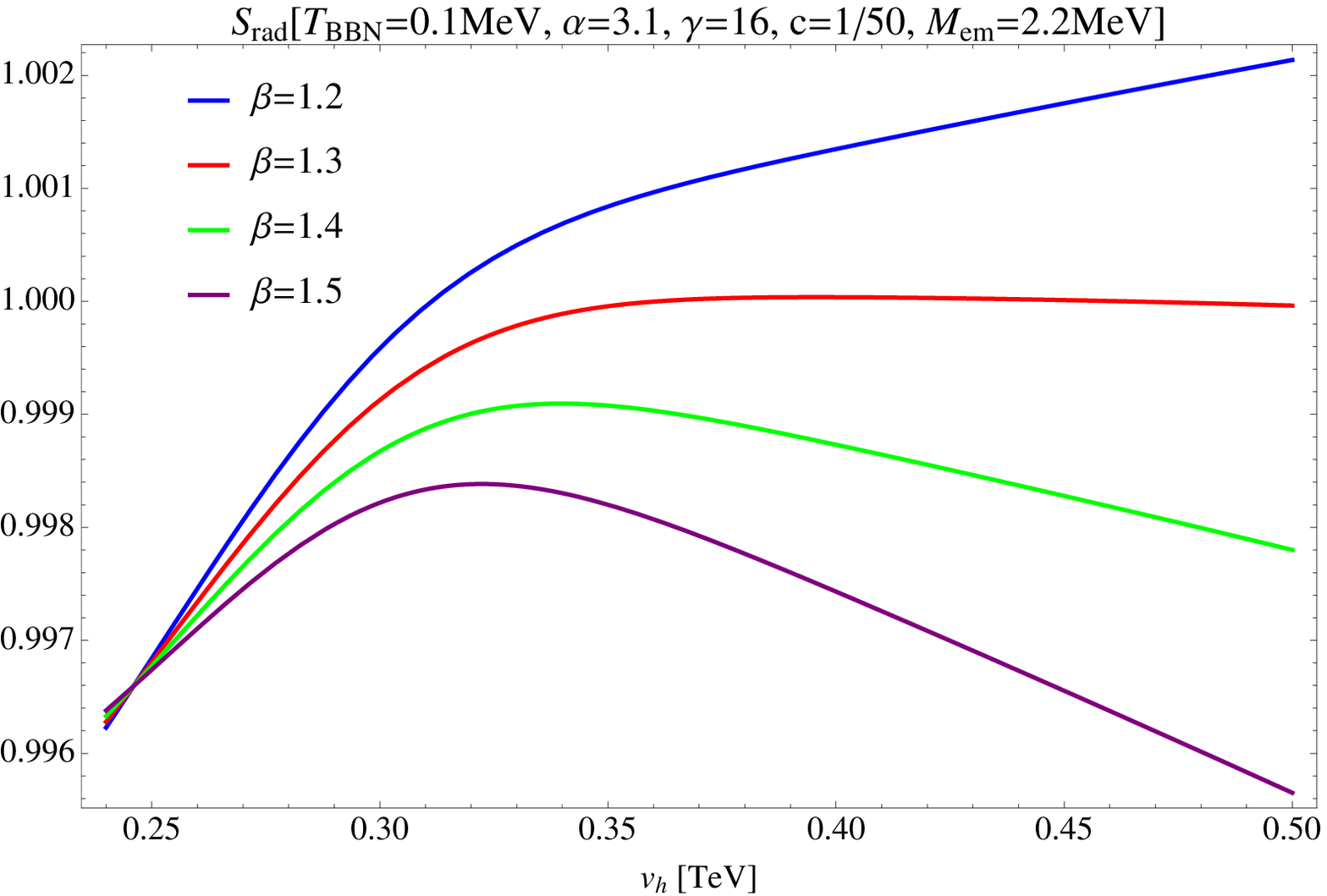}
\end{center}
\end{minipage}
\\
\\
\begin{minipage}{0.5\hsize}
\begin{center}
\includegraphics[width=8.5cm]{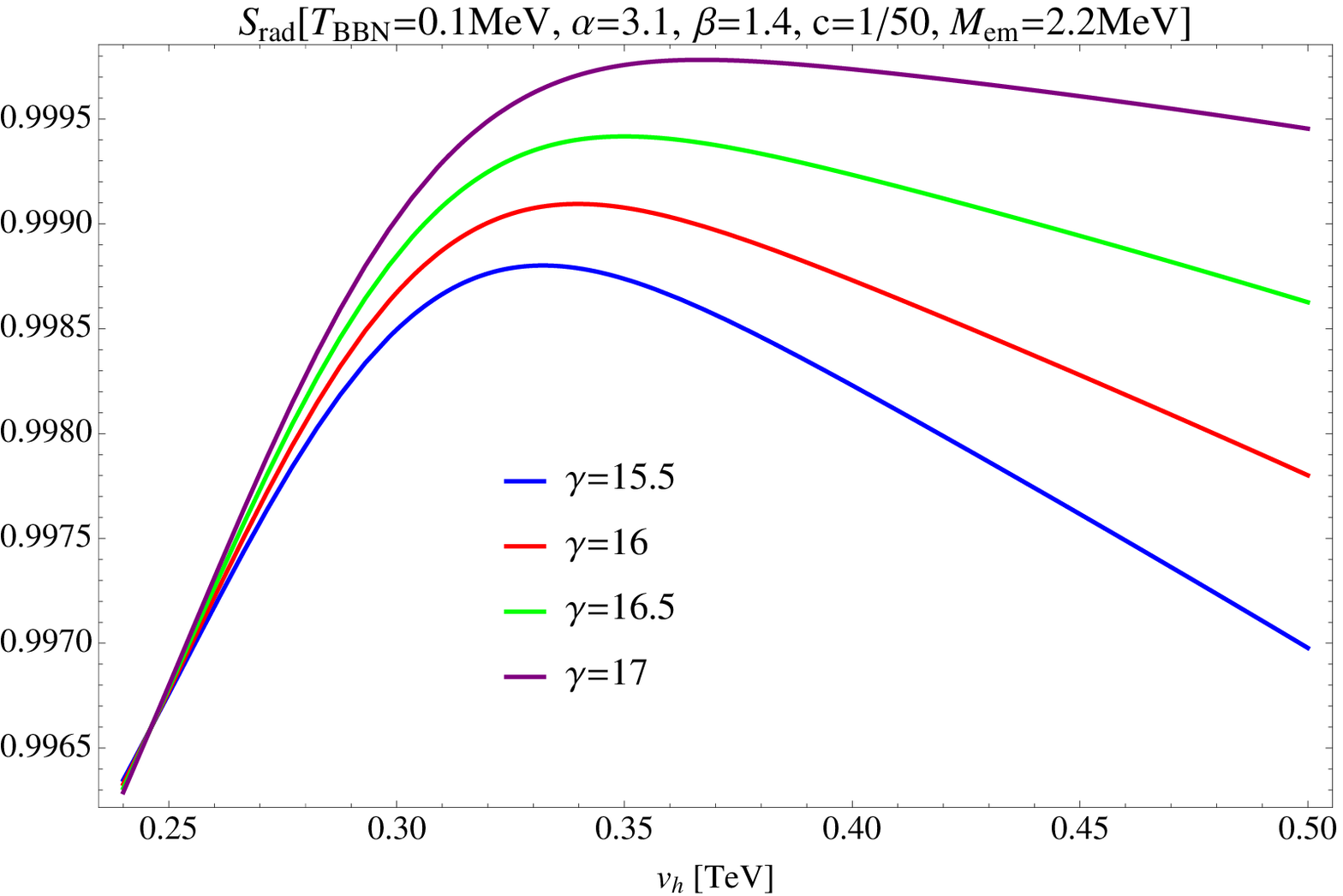}
\end{center}
\end{minipage}
\end{tabular}
\end{center}
\caption{The parameter dependences of $S_{rad}$ as a function of $v_{h}$. Here, $T_{BBN}$ and $M_{em}$ are fixed respectively to $0.1$ MeV and $2.2$ MeV, and $c$ is chosen to $1/50$.  In the upper left, upper right and lower figures, $\alpha,\beta $ and $\gamma$ are changed respectively.}
\label{fig:radiation1}
\end{figure}

We can also check that the quantitative behavior of $S_{rad}$ does not change even if $\alpha$, $\beta$ and $\gamma$ are varied by ${\cal{O}}(0.1)$. See Fig.\ref{fig:radiation1}. 
It is interesting to see whether QCD predicts the such values of $\alpha$, $\beta$ and $\gamma$ that $S_{rad}$ becomes maximum at $v_{h}=246$GeV precisely. \\

\begin{figure}[t]
\begin{center}
\includegraphics[width=10cm]{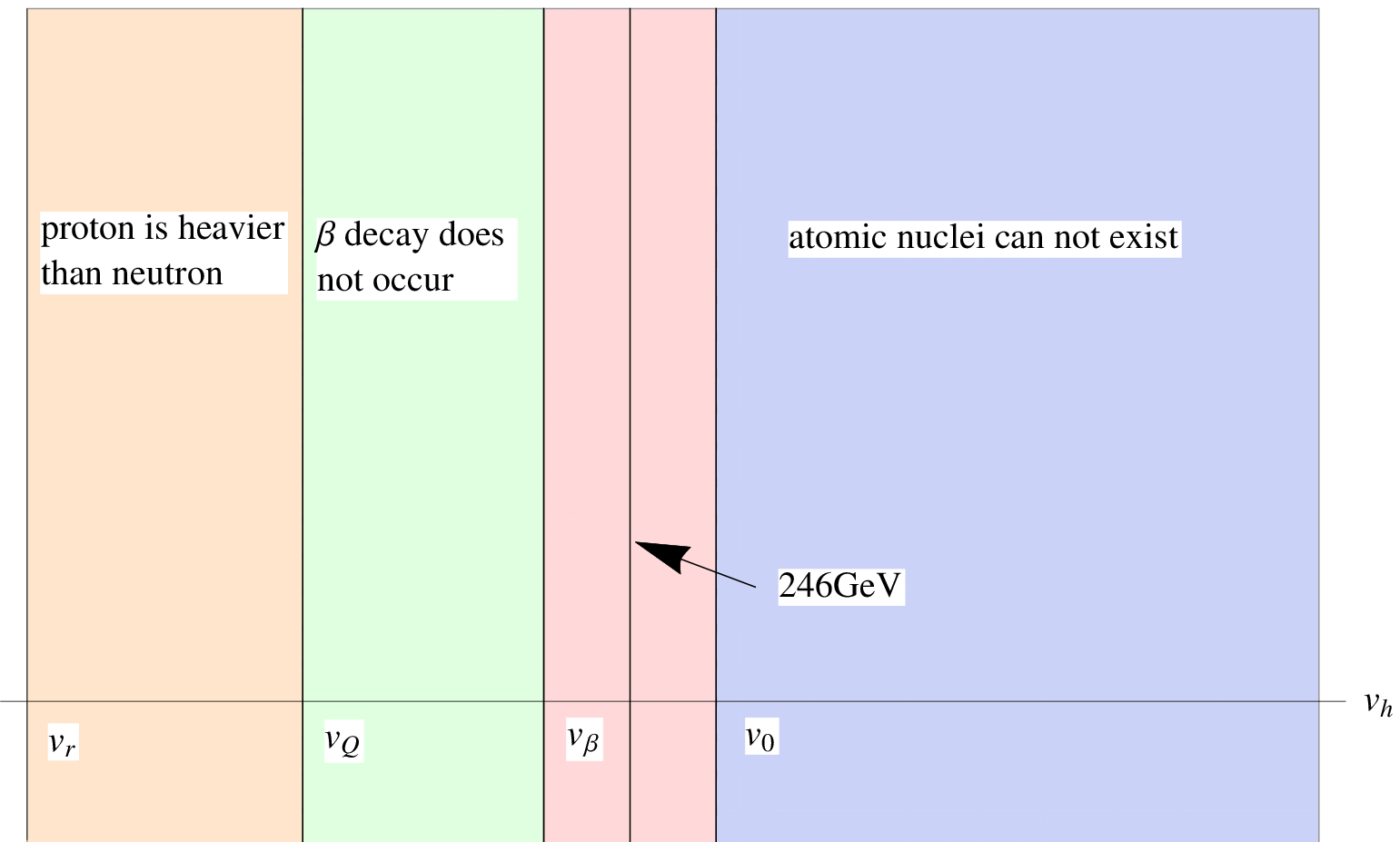}
\caption{How physics changes when the Higgs expectation value $v_{h}$ is varied. } 
\label{fig:vhregion}
\end{center}
\end{figure}

A few comments are in order. Originally, the $c\cdot X_{n}$ term in $S_{rad}$ comes from the existence of helium nuclei, which is guaranteed by
\be m_{\text{He}}=2(m_{p}+m_{n})-\Delta<4m_{p}+2m_{e},\e
where $\Delta$ is the binding energy of a helium nucleus. This is equivalent to 
\be 2(Q-m_{e})<\Delta,\e
However, when we change $v_{h}$ with the Yukawa couplings   fixed at the observed values, $\Delta$ and $2(Q-m_{e})$ can become equal at some point $v_{h}=v_{0}>246$GeV. This is  because $Q-m_{e}$ is an increasing function of $v_{h}$ and $\Delta$ is expected to be a decreasing function of $v_{h}$ \footnote{ The nucleon-nucleon interaction comes from the pion exchange, which becomes weak if the pion mass $m_{\pi}$ becomes large, thus $\Delta$ is a decreasing function of $v_{h}$ because $m_{\pi}$ is an increasing function of $v_{h}$ (see Eq.(\ref{eq:hadronmass})).}. Therefore, in the $v_{h}>v_{0}$ region, atomic nuclei can not exist (see Fig.\ref{fig:vhregion}). However, a simple analysis indicates that $v_{0}$ is greater than $1$TeV, where $X_{n}$ is very close to zero (see Fig.\ref{fig:xnvalue}). Therefore, we can trust the so far analysis even in this region\footnote{ Although the  $v_{h}$ dependence of $\Delta$ is complicated, the upper bound of $v_{0}$ can be obtained by
$Q-m_{e}=\Delta(v_{h}=246\text{GeV})/2=14\text{MeV}.\label{eq:vhmax}$
For example, if we assume $\beta=1.4$ and use $(m_{u},m_{d})=(2.3\text{MeV},4.8\text{MeV})$ \cite{Beringer:1900zz}, $M_{em}$ becomes $2.2$MeV, and $v_{0}^{(\text{Max})}$ is given by
$ v_{0}^{(\text{Max})}(\beta=1.4,M_{em}=2.2\text{MeV})=1.3\text{TeV}.\nonumber\label{eq:limit1}$}.

On the other hand, in the $v_{h}<246$GeV region, there are three critical points where  $Q+m_{e},Q$ and $Q-m_{e}$ becomes zero, which we denote respectively by
\be v_{r}\h{2mm},\h{2mm}v_{Q}\h{2mm},\h{2mm}v_{\beta}.\e
In the region $v_{Q}<v_{h}<v_{\beta}$, the beta decay does not occur. This case is already included in the so far analysis. In the $v<v_{Q}$ region, proton becomes heavier than neutron. Even if such universe exists, there should also be atomic nuclei because of the isospin-symmetry. In the $v_{h}<v_{r}$ region, protons becomes unstable, and again atomic nuclei cease to exist if $v_{h}$ is much smaller than $v_{r}$.  As a result, $S_{rad}$ increases as $v_{h}$ gets smaller.  It is interesting to see whether the peak around $v_{h}={\cal{O}}(300\text{GeV})$ is the global maximum or not. 

\section{Summary and Discussion}\label{sec:summary}
\begin{figure}
\begin{center}
\includegraphics[width=9cm]{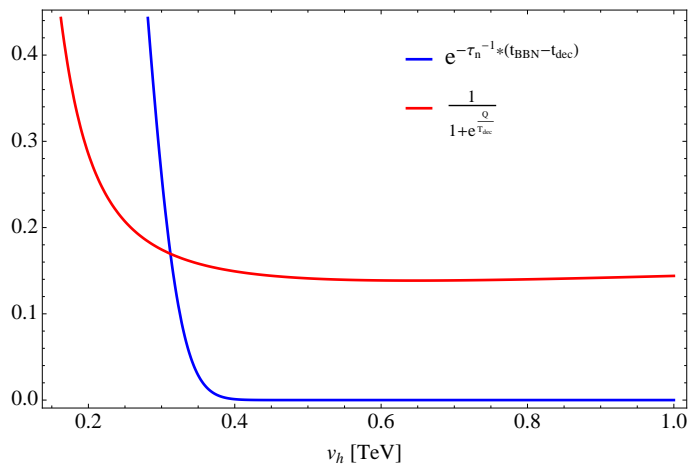}
\end{center}
\caption{The $v_{h}$ dependence of the two factors of $X_{n}$. Here, we show the case where $\beta=1.4$ and $M_{em}=2.2$MeV. Blue and Red lines show $\exp\left(-\tau_{n}^{-1}(t_{BBN}-t_{dec})\right)$ and $1/\left(1+\exp(Q/T_{dec})\right)$, respectively.}
\label{fig:factors}
\end{figure}

In this paper, we have shown that the radiation of the universe $S_{rad}$ at the final stage has a maximum around $v_{h}={\cal{O}}(300\text{GeV})$ when the Higgs self coupling, the gauge couplings and the Yukawa couplings are fixed. The $v_{h}$ dependence of $S_{rad}$ is given by Eq.(\ref{eq:apprad}). We have seen that $m_{p}^{\frac{4}{3}}\times\tau_{p}^{\frac{2}{3}}$ is a decreasing function for typical values of $\alpha$, $\beta$ and $\gamma$ while $1-cX_{n}$ increases rather rapidly for smaller values of $v_{h}$ and becomes constant one around $v_{h}=300\text{GeV}$. Therefore, $S_{rad}$ becomes maximum at the scale where $1-cX_{n}$ becomes one. 
As we have seen in subsection \ref{subsec:Xn}, $X_{n}$ is the product of
\be \frac{1}{1+\exp\left(\frac{Q}{T_{dec}}\right)},\label{eq:Qterm}\e
and
\be \exp\left(-\tau_{n}^{-1}(t_{BBN}-t_{dec})\right).\label{eq:betaterm}\e
Actually Eq.(\ref{eq:betaterm}) determines where $X_{n}$ becomes zero (see Fig.\ref{fig:factors}) and thus $1-cX_{n}$ becomes one. As a result, the maximum point of $S_{rad}$ is qualitatively given by
\be \tau_{n}^{-1}\times t_{BBN}\simeq1,\e
from which we obtain
\be v_{h}\simeq\frac{T_{BBN}^{2}}{M_{pl}y_{e}^{5}}.\e
This shows a surprising and mysterious relation between $v_{h}$, $T_{BBN}$, $y_{e}$ and $M_{pl}$. 

In conclusion, we mention the possibilities that we can make further predictions by combining the maximum entropy principle with the other principle such as the stability of the Higgs potential. For example, we can consider the U(1) gauge coupling $g_{Y}$. As seen in Fig.\ref{fig:radiation}, $S_{rad}$ increases if $M_{em}$ is decreased. Because $M_{em}$ mainly depends on the $g_{Y}$, the smaller values of $g_{Y}$ are favored by the maximum entropy principle. Thus, if there exists a lower bound of $g_{Y}$, we can conclude that $g_{Y}$ is fixed to that value. From the recent analyses \cite{quartic} based on the observed Higgs mass, it is possible that the Higgs potential is marginally stable up to the energy scale $10^{17}\sim10^{18}$GeV. In other words, if $g_{Y}$ is smaller than the observed value, the Higgs potential is unstable. This means that the present value of $g_{Y}$ is at the lower bound, which is consistent  with the above argument. It would be interesting to consider various use of the maximum entropy principle. 

\section*{Acknowledgement} 
The work of Y.~H. is supported by a Grant-in-Aid for Japan Society for the Promotion of  Science(JSPS) Fellows No.25$\cdot$1107.

\appendix 
\def\thesection{Appendix \Alph{section}}

\section{Including the $v_{h}$ Dependence of $T_{BBN}$}\label{app:bd}
In this appendix, we take the $v_{h}$ dependence of $T_{BBN}$ into account. $T_{BBN}$ is the temperature where the ratio of deuterons to all nucleons
\be X_{d}:=\frac{n_{d}}{n_{N}}\e
becomes ${\cal{O}}(1)$. In the following,  we find
\be T_{BBN}\simeq 0.03\times B_{d},\e
where $B_{d}$ is the deuteron binding energy $B_{d}$. 

In thermal equilibrium, the number density of the particle species $i$ having the heavy mass $m_{i}\gg T$ is given by 
\be n_{i}=g_{i}\left(\frac{m_{i}T}{2\pi}\right)^{\frac{3}{2}}\times\exp\left(-\frac{m_{i}-\mu_{i}}{T}\right),\label{eq:density}\e 
where $g_{i}$ is the internal degrees of freedom, and $\mu_{i}$ is the chemical potential. If $T\ll 1$GeV, we can use Eq.(\ref{eq:density}) for protons, neutrons and deuterons. Therefore, we obtain 
\begin{align} \frac{n_{d}}{n_{p}n_{n}}&=\frac{3(2\pi)^{3/2}}{4}\times\left(\frac{m_{d}}{m_{p}m_{n}T}\right)^{3/2}\exp\left(-\frac{m_{d}-m_{p}-m_{n}}{T}\right)\nonumber\\
&=\frac{3(2\pi)^{3/2}}{4}\times\left(\frac{m_{d}}{m_{p}m_{n}T}\right)^{3/2}\times\exp\left(\frac{B_{d}}{T}\right):=f(T,B_{d}),\end{align}
where we have used the relation between the chemical potentials\footnote{Protons, neutrons and deuterons are in thermal equilibrium through $p+n\leftrightarrow d+\gamma$.}
\be \mu_{d}=\mu_{p}+\mu_{n}.\e
By using 
\be n_{p}=(1-X_{n})n_{N}\h{2mm},\h{2mm}n_{n}=X_{n}n_{N},\e
and the baryon to photon ratio
\be \frac{n_{\gamma}}{n_{N}}\simeq6\times10^{10},\e
$X_{d}$ is given by
\begin{align} X_{d}&=\frac{n_{d}}{n_{N}}=\frac{n_{d}}{n_{p}n_{n}}\times\frac{n_{p}n_{n}}{n_{N}}\nonumber\\
&=f(T,B_{d})\times X_{n}(1-X_{n})n_{N}\nonumber\\
&=f(T,B_{d})\times X_{n}(1-X_{n})\times\frac{10^{-10}}{6}\times n_{\gamma}\nonumber\\
&\simeq1.35\times10^{-10}\times \left(\frac{T}{1\text{GeV}}\right)^{3/2}\exp\left(\frac{B_{d}}{T}\right).\end{align}
Here, we have put $X_{n}(1-X_{n})\simeq0.1$. This will be verified by seeing that $X_{n}$ has an ${\cal{O}}(0.1)$ value as a function of $v_{h}$ around $v_{h}=300$GeV (see Fig.\ref{fig:xnvalue}). By solving $X_{d}=1$ numerically, we can obtain 
\be T_{\text{BBN}}\simeq 0.03\times B_{d}.\label{eq:Tbinding}\e
\begin{figure}
\begin{minipage}{0.5\hsize}
\begin{center}
\includegraphics[width=8.5cm]{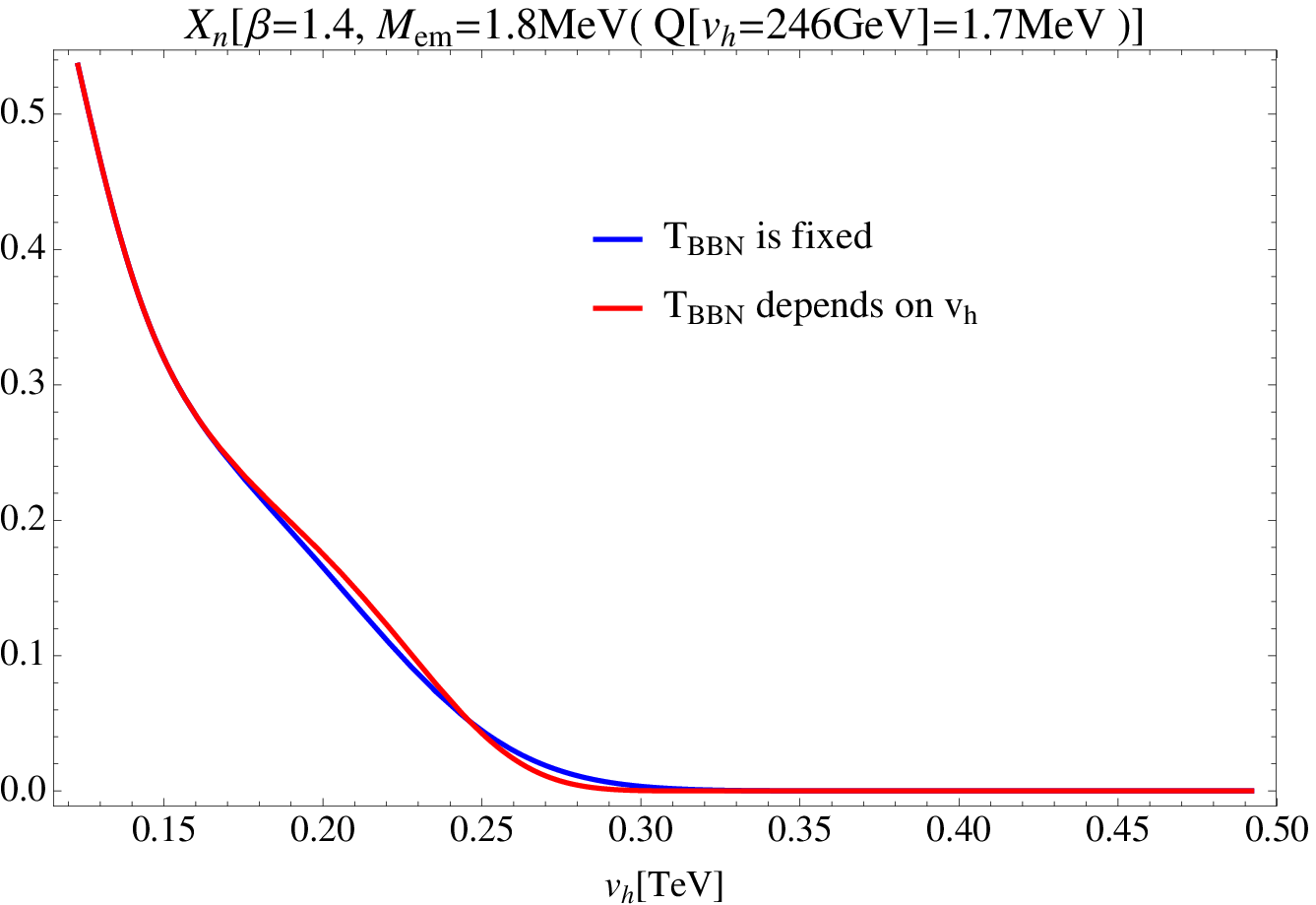}
\end{center}
\end{minipage}
\begin{minipage}{0.5\hsize}
\begin{center}
\includegraphics[width=8.5cm]{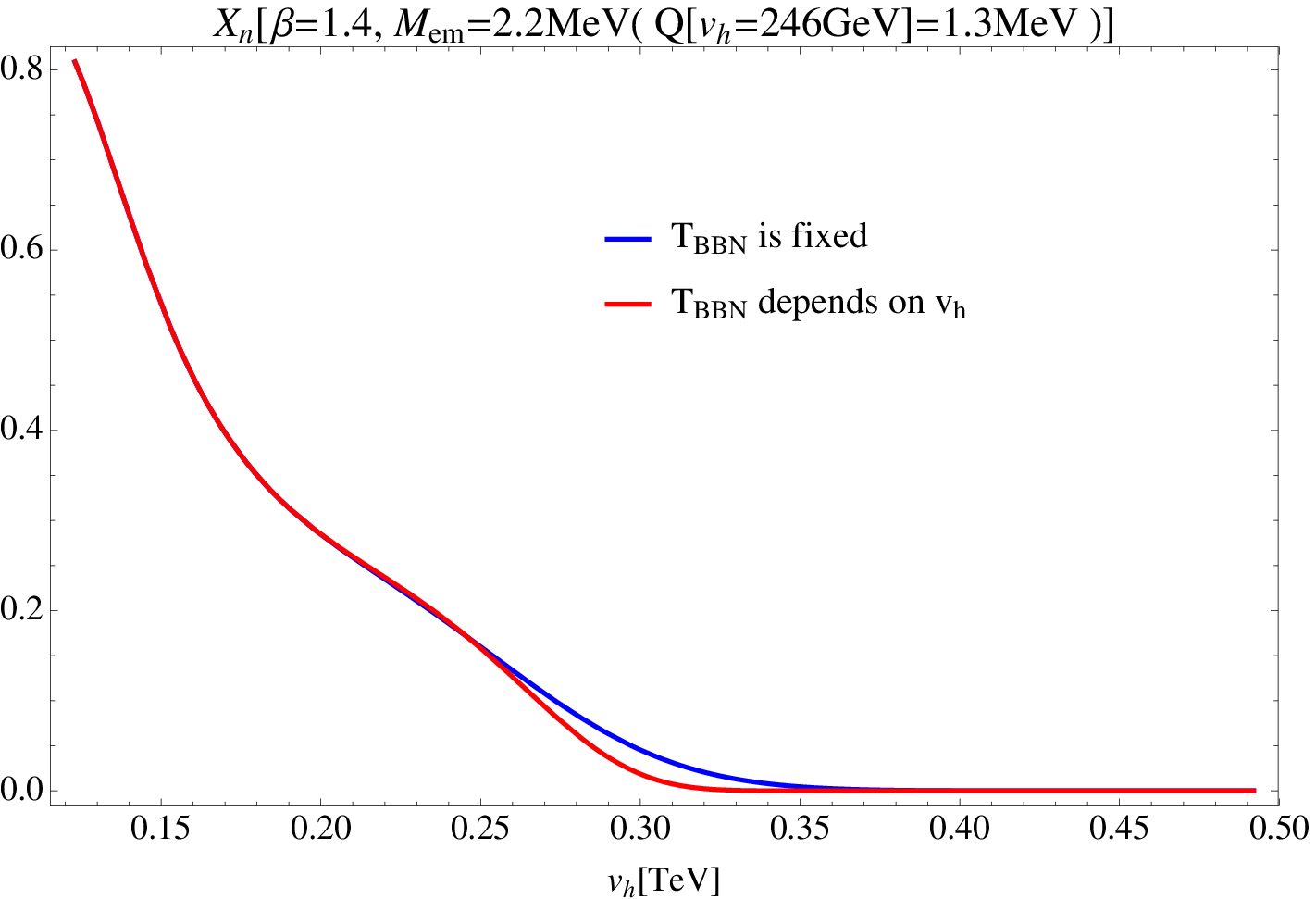}
\end{center}
\end{minipage}
\caption{$X_{n}$ as a function of $v_{h}$ in the cases where the $v_{h}$ dependence of $T_{BBN}$ is included (Red), and where $T_{BBN}$ is fixed to that of $v_{h}=246$GeV (Blue).  In the left and right panels, we assume $M_{em}=1.8$MeV and $M_{em}=2.2$MeV, respectively.} 
\label{fig:XnBBN}
\end{figure}
Although determining $B_{d}$ as a function of $v_{h}$ is difficult, the qualitative behavior is easily understood. Because  $m_{\pi}^{2}\propto v_{h}$, the strength of the nuclear force is roughly given by $v_{h}^{-1}$. Therefore, $B_{d}$ should be a decreasing function of $v_{h}$. Here, we use the recent result \cite{Berengut:2013nh}\footnote{Although, in \cite{Berengut:2013nh}, $B_{d}$ is calculated in the $0.5<m_{q}/m^{(\text{phy})}_{q}<2$ or $123\text{GeV}<v_{h}<492$GeV region, where the chiral perturbation theory seems to be reliable, this region is enough to examine the maximum point of $S_{rad}$. } in which $B_{d}$ is calculated as a function of the current quark masses by using the effective chiral perturbation theory. The result in \cite{Berengut:2013nh} agrees with the above heuristic argument. 

Now we consider $X_{n}$ when we take the $v_{h}$ dependence of $T_{BBN}$ into consideration. First we read off  $B_{d}$ as a function of $v_{h}$ from \cite{Berengut:2013nh}, then by using  Eq.(\ref{eq:Tbinding}), Eq.(\ref{eq:expansion}) and Eq.(\ref{eq:Xnvalue}), we obtain $X_{n}$ as a function of $v_{h}$. See Fig.\ref{fig:XnBBN}. \footnote{One can see that the $v_{h}$ dependence of $T_{BBN}$ has an effect to decrease (increase) $X_{n}$ in the $v_{h}> (<) 246$GeV region. This can be understood intuitively; as discussed above, because $B_{d}$ is a decreasing function of $v_{h}$, $T_{BBN}$ is also a decreasing function. Thus, if $v_{h}$ is large, $T_{BBN}$ becomes small, and the time when the BBN starts becomes large. As a result, the beta decay lasts for a longer time, and $X_{n}$ decreases.}

\begin{figure}
\begin{center}
\begin{tabular}{c}
\begin{minipage}{0.5\hsize}
\begin{center}
\includegraphics[width=9cm]{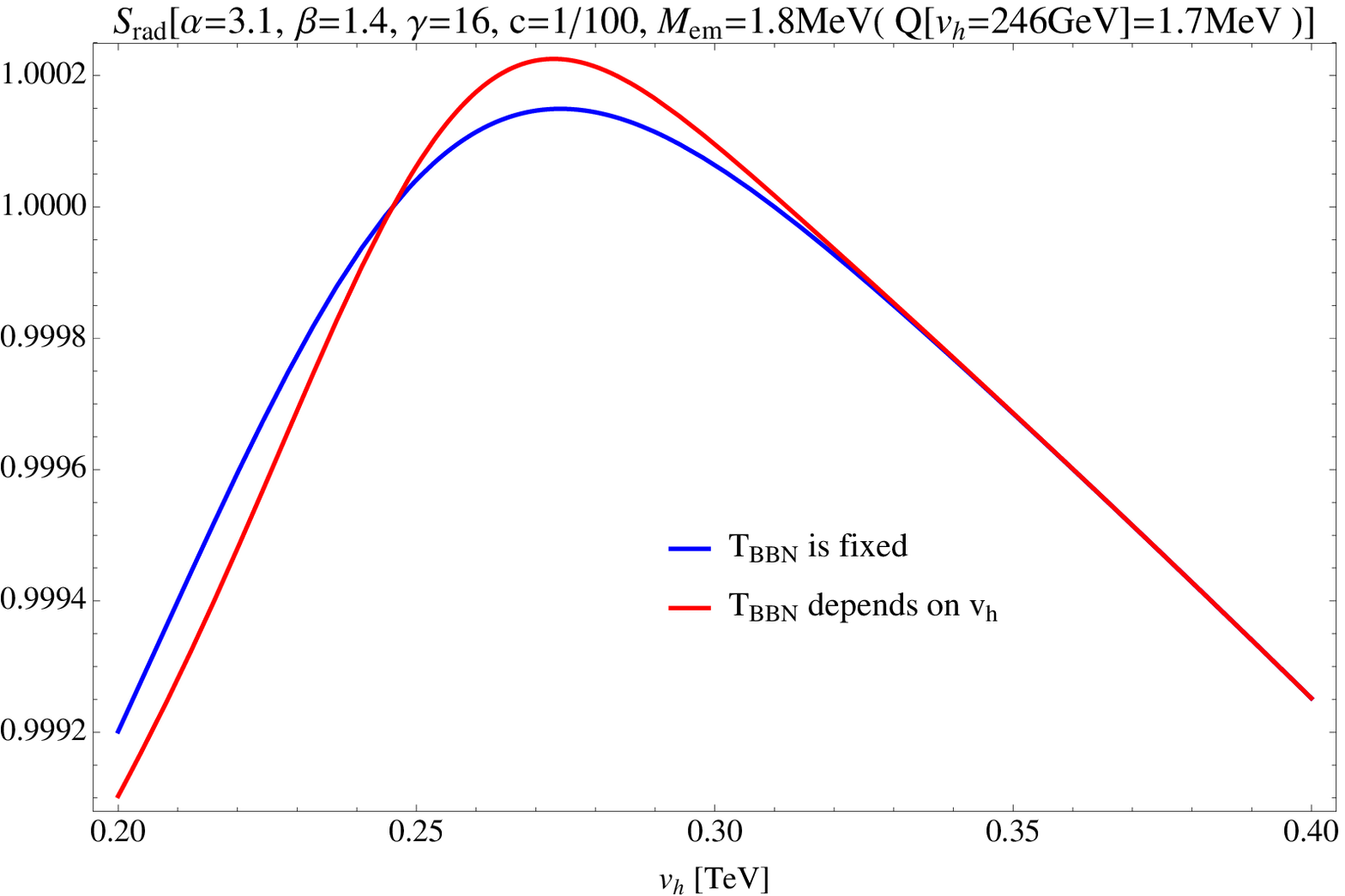}
\end{center}
\end{minipage}
\begin{minipage}{0.5\hsize}
\begin{center}
\includegraphics[width=9cm]{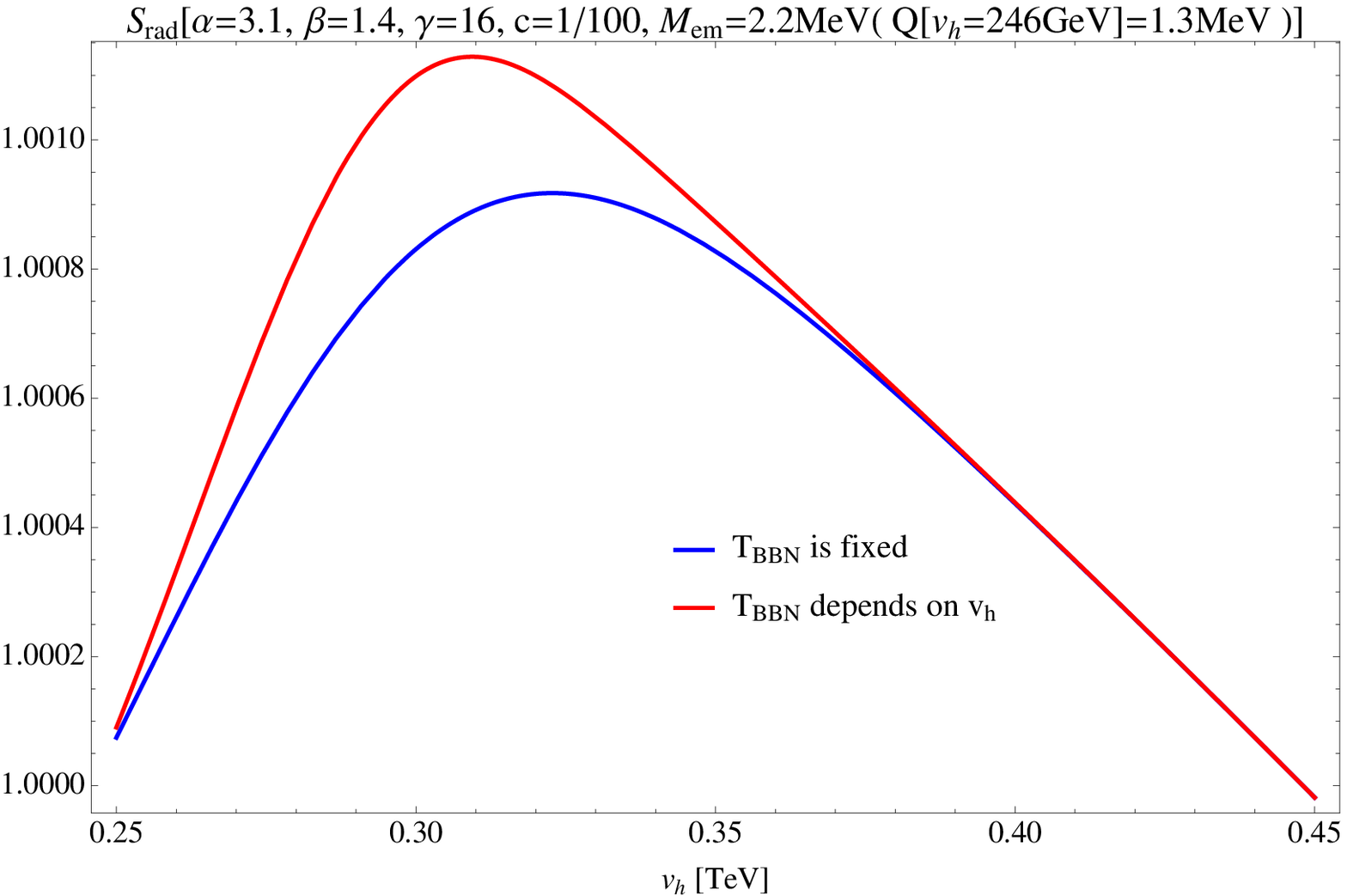}
\end{center}
\end{minipage}
\end{tabular}
\end{center}
\caption{$S_{rad}$ as a function of $v_{h}$ in the cases where the $v_{h}$ dependence of $T_{BBN}$ is included (Red), and where $T_{BBN}$ is fixed to that of $v_{h}=246$GeV (Blue). Here, we have chosen $c=1/100$. In the left and right panels, we assume $M_{em}=1.8$MeV and $M_{em}=2.2$MeV, respectively.}
\label{fig:radiation2}
\end{figure}

By using Eq.(\ref{eq:apprad}), we obtain $S_{rad}$ as a function of $v_{h}$. See Fig.\ref{fig:radiation2} \footnote{Qualitatively, the radiation increases (decreases) in the  $246\text{GeV}<v_{h}\h{1mm}(246\text{GeV}>v_{h})$ region because $X_{n}$ decreases (increases) in this region.}. We can see that the maximum of $S_{rad}$ slightly changes.  

\section{ The Proton Life Time}\label{app:proton}
In this Appendix, we give the formula for the proton life time. We start from the effective Lagrangian \cite{Claudson:1981gh}. The process 
\be p\rightarrow e^{+}+\pi_{0}\label{eq:exa1}\e
is described by
\be  {\cal{L}}_{decay}=i\frac{m_{p}^{2}}{M_{G}^{2}}\left(c_{1}\pi_{0}\bar{e}^{c}p+c_{2}\pi_{0}\bar{e}^{c}\gamma_{5}p\right)+\text{h.c},\e
where $M_{G}$ is the GUT scale. Although $c_{1}$ and $c_{2}$ may depend on the current quark masses due to the wave function of a proton and pion, we assume that they are constants in this paper.

Then, the partial width of Eq.(\ref{eq:exa1}) is given by
\begin{align} \Gamma_{p}&=\frac{1}{2m_{p}}\int\frac{d^{3}p^{'}}{(2\pi)^{3}2p^{'}_{0}}\int\frac{d^{3}k}{(2\pi)^{3}2k_{0}}\frac{1}{2}\cdot \sum_{spin} |{\cal{M}}|^{2}\h{2mm}(2\pi)^{4}\delta(p-p^{'}-k)\nonumber\\
&=\frac{m_{p}^{5}(c_{1}^{2}+c_{2}^{2})}{2^{4}\pi M_{G}^{4}}\times\left(1-\left(\frac{m_{\pi}}{m_{p}}\right)^{2}\right)^{2}.
\end{align}
Taking into account the other process
\be p\rightarrow \pi^{+}+\bar{\nu} \label{eq:exa2},\e
we have the approximate expression for $G$ in Eq.(\ref{eq:proton life})
\be G=constant\times\left(1-\frac{m_{\pi}^{2}}{m_{p}^{2}}\right)^{2}.
\label{eq:appG}\e 
Let us calculate $\eta$ in Eq.(\ref{eq:defeta}) which has been explained in Section\ref{sec:para}. We neglect the electron mass in the following argument. Because $c_{1}$ and $c_{2}$ are constants in our assumption, $G$ becomes
when $m_{u,d}\ll m_{p}$. Thus, by using Eq.(\ref{eq:hadronmass}) and Eq.(\ref{eq:appG}), we have
\begin{align} m_{p}^{\frac{4}{3}}\times\tau_{p}^{\frac{2}{3}}&\propto\frac{1}{m_{p}^{2}}G^{-\frac{2}{3}}\nonumber\\
&\propto\left(1+\frac{\beta(2m_{u}+m_{d})}{\alpha\Lambda_{QCD}}\right)^{-2}\times\left(1-\frac{\gamma(m_{u}+m_{d})}{2\alpha^{2}\Lambda_{QCD}}\right)^{-\frac{4}{3}}\nonumber\\
&\simeq\left(1-\frac{2\beta(2m_{u}+m_{d})}{\alpha\Lambda_{QCD}}+\frac{2\gamma(m_{u}+m_{d})}{3\alpha^{2}\Lambda_{QCD}}\right)\nonumber\\
&=1-\left(\frac{4\beta}{\alpha}-\frac{2\gamma}{3\alpha^{2}}\right)\cdot\frac{m_{u}}{\Lambda_{QCD}}-\left(\frac{2\beta}{\alpha}-\frac{2\gamma}{3\alpha^{2}}\right)\cdot\frac{m_{d}}{\Lambda_{QCD}}.\label{eq:eta1}\end{align}
For example, if we choose 
\be \alpha=3.1\h{2.5mm},\h{2.5mm}\beta=1.4\h{2.5mm},\h{2.5mm}\gamma=16,\e
Eq.(\ref{eq:eta1}) becomes
\be m_{p}^{\frac{4}{3}}\times\tau_{p}^{\frac{2}{3}}\propto1-0.69\cdot\frac{m_{u}}{\Lambda_{QCD}}+0.20\cdot\frac{m_{d}}{\Lambda_{QCD}}.\e
Thus, if we use the typical values $(m_{u},m_{d})=(2.3\text{MeV},4.8\text{MeV})$ \cite{Beringer:1900zz}, the coefficient of $v_{h}$ becomes negative, which means that $m_{p}^{\frac{4}{3}}\times\tau_{p}^{\frac{2}{3}}$ is the monotonically decreasing function for small $v_{h}$.

\section{Parameter Region where $c>0.01$}\label{app:c}
\begin{figure}[!h]
\begin{center}
\begin{tabular}{c}
\begin{minipage}{0.5\hsize}
\begin{center}
\includegraphics[width=6cm]{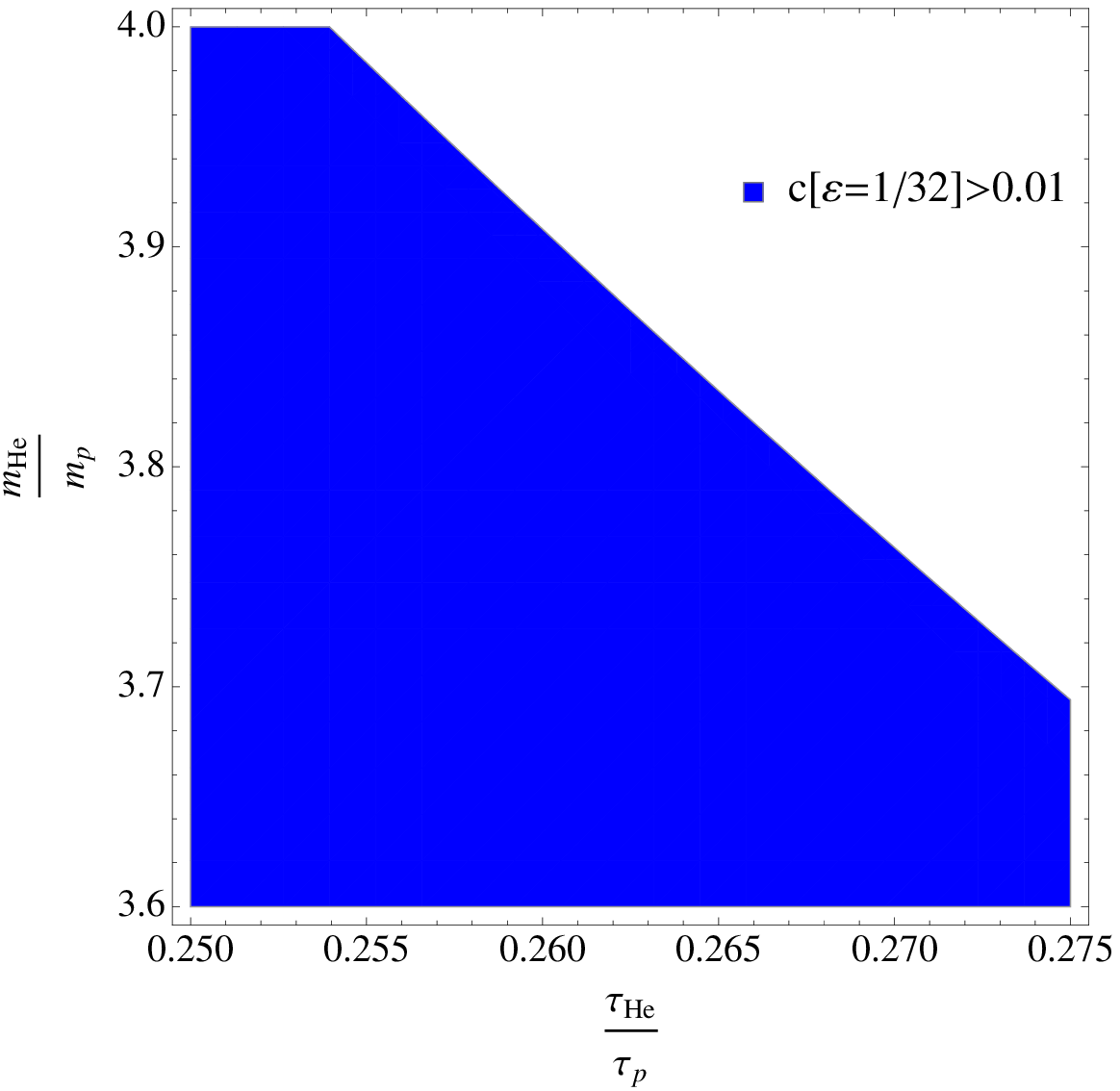}
\end{center}
\end{minipage}
\begin{minipage}{0.5\hsize}
\begin{center}
\includegraphics[width=6cm]{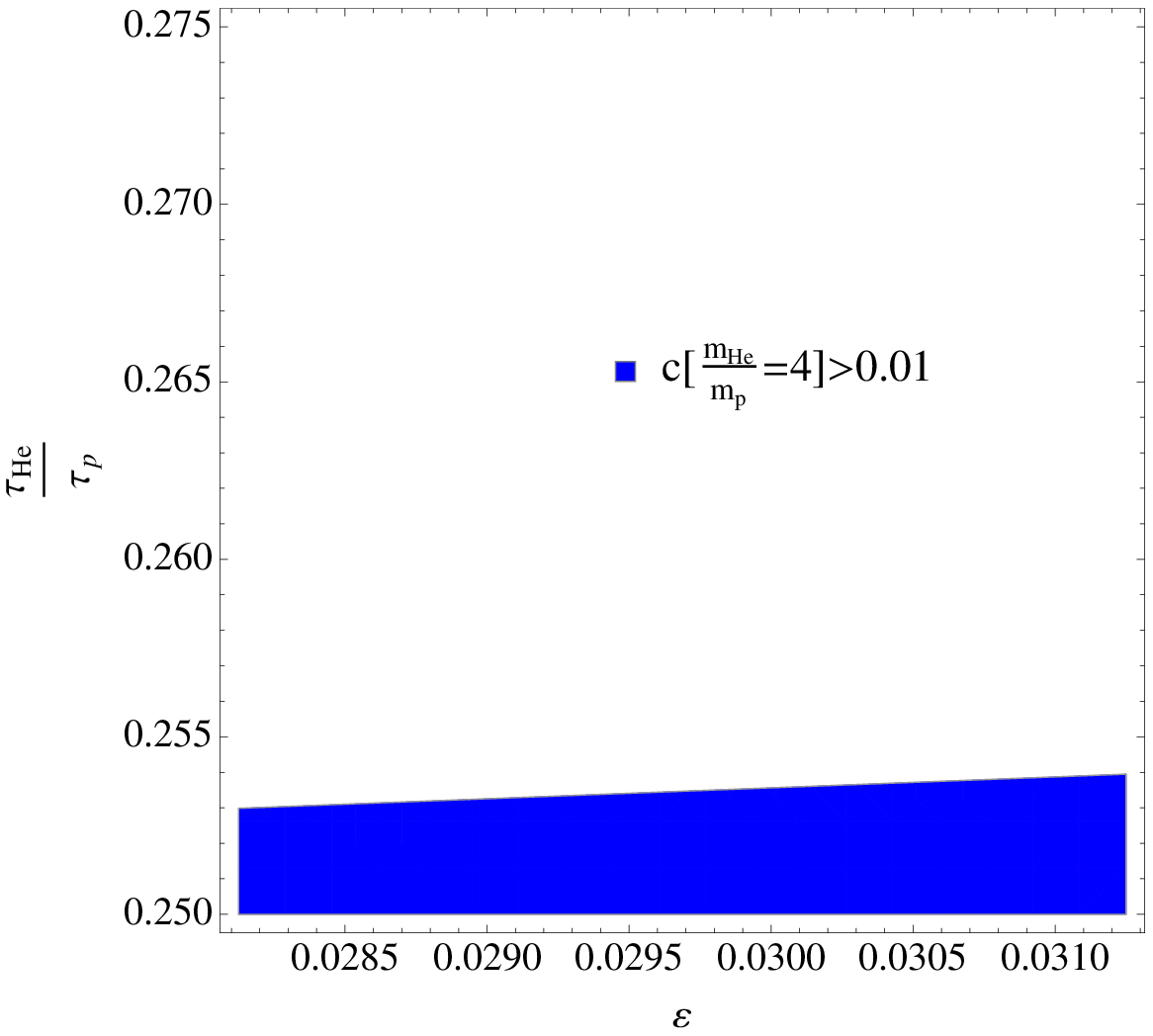}
\end{center}
\end{minipage}
\\
\begin{minipage}{0.5\hsize}
\begin{center}
\includegraphics[width=7.5cm]{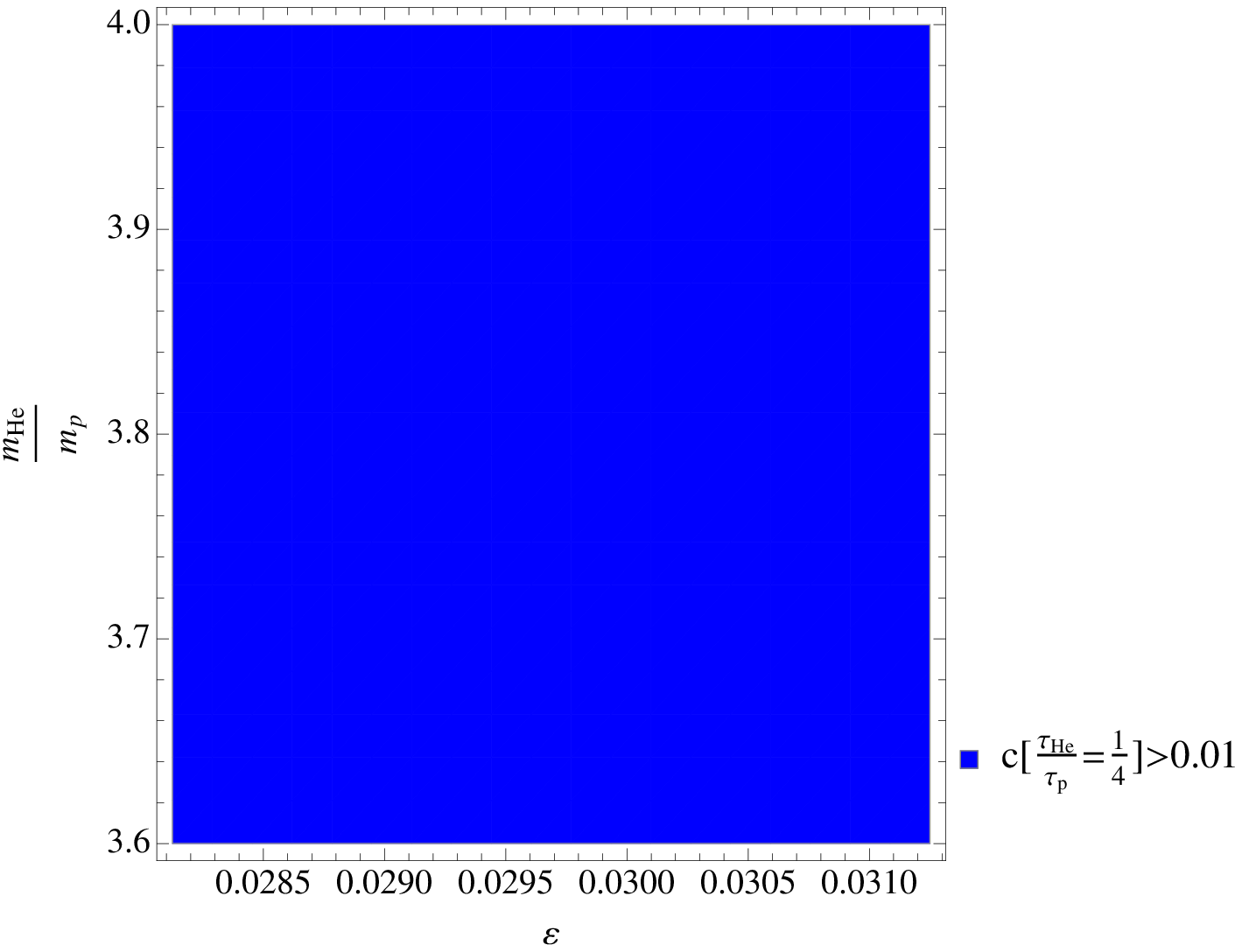}
\end{center}
\end{minipage}
\end{tabular}
\end{center}
\label{fig:cregion2}
\caption{The parameter regions where $c>0.01$ when we allow $\frac{m_{He}}{m_{p}}$, $\frac{\tau_{He}}{\tau_{p}}$ and $\epsilon$ to change by $10\%$ from their natural values. One can see that only in the region where $\tau_{He}$ is large, $c$ can be smaller than $0.01$. } 
\end{figure}

One of the crucial assumptions in our argument is having fixed $c$ to a small positive value such as $1/50$ or $1/100$. By solving the differential equations (\ref{eq:Npdiff})-(\ref{eq: radiation2}), we can examine how naturally we have $c>0.01$ when we change $\frac{m_{He}}{m_{p}}$,$\frac{\tau_{He}}{\tau_{p}}$ and $\epsilon$ by $10\h{1mm}\%$ from their roughly estimated values $(4,\frac{1}{4},\frac{1}{32})$ \cite{Hamada:2014ofa}. In Fig.9, the $c>0.01$ region is shown by blue. One can see that a wide range of parameters gives $c>0.01$. However, the large $\tau_{He}$ region is not allowed, which comes from the fact that the life time of an atomic nucleon increases the radiation of the universe \footnote{Qualitatively this can be understood as follows; if the matter having the energy $\Delta M$ decays and becomes radiation, we can obtain $\frac{\Delta M}{a^{3}(t)}=\frac{\Delta S_{rad}}{a^{4}(t)}$, namely $\Delta S_{rad}=a(t)\times\Delta M$ from the conservation of the energy density. Thus, if the decay time becomes large, because $a(t)$ becomes large, the radiation increases.}. Thus, if the effect from $\tau_{He}$ dominates in the small $v_{h}$ region, $c$ becomes negative, and the peak of $S_{rad}$ obtained in Section\ref{sec:big} disappears. To make more quantitative argument, we need to understand how $\tau_{He}$ and $m_{He}$ depend on the quark masses.

\end{document}